# Improper Ferroelectricity at the Monolayer Limit

Yilin Evan Li[1*], Harikrishnan KP[2*], Haidong Lu[3*], Rachel A. Steinhardt[1,14*], Megan E. Holtz[1,2,15*], Mario Brützam[4], Matthew M. Dykes[5], Elke Arenholz[6], Sankalpa Hazra[7], Adriana LaVopa[1,16], Xiaoxi Huang[5], Wenwen Zhao[2], Piush Behera[8], Maya Ramesh[1], Evan Krysko[1], Venkatraman Gopalan[7,9], Ramamoorthy Ramesh[10,11], Craig J. Fennie[2], Robert J. Cava[13], Christo Guguschev[4], Alexei Gruverman[3], David A. Muller[2,12], Darrell G. Schlom[1,4,12]

[1] Department of Materials Science and Engineering, Cornell University, Ithaca, NY, USA.
[2] School of Applied and Engineering Physics, Cornell University, Ithaca, NY, USA.
[3] Department of Physics and Astronomy, University of Nebraska, Lincoln, NE, USA.
[4] Leibniz-Institut für Kristallzüchtung, Berlin, Germany.
[5] Department of Physics, Cornell University, Ithaca, NY, USA.
[6] Physical Sciences Division, Pacific Northwest National Laboratory, Richland, WA, USA.
[7] Department of Materials Science and Engineering, The Pennsylvania State University, University Park, PA, USA.
[8] Department of Materials Science and Engineering, University of California, Berkeley, Berkeley, CA, USA.
[9] Department of Physics, The Pennsylvania State University, University Park, PA, USA.
[10] Department of Materials Science and Nanoengineering, Rice University, Houston, TX, USA.
[11] Department of Physics and Astronomy, Rice University, Houston, TX, USA.
[12] Kavli Institute at Cornell for Nanoscale Science, Ithaca, NY, USA.
[13] Department of Chemistry, Princeton University, Princeton, NJ, USA.
[14] Present address: Components Research, Intel Corporation, Hillsboro, OR, USA.
[15] Present address: Department of Metallurgical and Materials Engineering, Colorado School of Mines, Golden, CO, USA.
[16] Present address: Department of Materials Science and Engineering, University of Florida, Gainesville, FL, USA.

* These authors contributed equally to this work

Ultrathin ferroelectric films with out-of-plane polarization and high Curie temperatures are key to miniaturizing electronic devices[1,2]. Most ferroelectrics employed in devices are proper ferroelectrics, where spontaneous polarization is the primary order parameter[3,4]. Unfortunately, the Curie temperature of proper ferroelectrics is drastically reduced as the ferroelectric becomes thin[5,6]; nearly all proper ferroelectrics need to be thicker than several unit cells[5,7–9]. The absence of an ultrathin limit has been predicted[10,11], but not verified[12,13] for improper ferroelectrics. These are ferroelectrics where the polarization emerges secondary to the primary order parameter, such as a structural distortion[14,15]. Here we report improper ferroelectricity with an undiminished Curie temperature in a 0.75-unit-cell-thick hexagonal $LuFeO_3$ ($h$-$LuFeO_3$) film grown on a $SrCo_2Ru_4O_{11}$ bottom electrode with an atomically engineered monolayer bridging layer. Our results demonstrate the absence of a critical thickness for improper ferroelectricity and provide a methodology for creating ultrathin improper ferroelectrics by stabilizing their primary order parameters.

Recent studies have showcased out-of-plane polarization in ultrathin $BiFeO_3$[16,17] and $ZrO_2$[18] films, even down to one-unit-cell thickness. Yet, this is not the case for most other proper ferroelectric thin films with out-of-plane polarization. As these proper ferroelectrics get thinner, they form smaller 180° in-plane domains, suffer from decreased Curie temperatures and polarization, and can eventually lose their polarization entirely at the two-dimensional limit[5–9]. This loss is mainly due to the depolarization field. In contrast, improper ferroelectrics follow a different mechanism. Their polarization couples linearly to some non-polar primary order parameter. On strong theoretical grounds, this dependence on the primary order parameter will make improper ferroelectricity more immune to depolarization fields because the electric polarization is preserved as long as the primary order parameter remains stable. Supporting this concept, density functional theory studies have suggested that single-domain improper ferroelectricity with out-of-plane polarization can persist in ultrathin films without a critical thickness, even when the depolarization field is unscreened[10,11].

This predicted robustness is, however, not seen experimentally in studies of model improper ferroelectrics including hexagonal $YMnO_3$ ($h$-$YMnO_3$)[12] and hexagonal $ScFeO_3$ ($h$-$ScFeO_3$)[13]. In both cases, a non-polar, trimer distortion that triples the unit cell serves as the primary order parameter[15]. These $h$-$ABO_3$ films appear "clamped" in the untrimerized structure near the interface with the substrates used, thwarting the needed primary structural distortion, and resulting in a loss of the secondary electric polarization until the films exceed a thickness of 2-unit cells (4 formula units)[12,13]. Consequently, we seek to eliminate substrate clamping by means of interface engineering in an attempt to push improper ferroelectricity to the monolayer limit. Our goal is to demonstrate that paraelectric-clamping is not an intrinsic barrier to synthesizing the desired improper ferroelectric polymorph on a substrate with a different crystal structure.

Our approach involves engineering an epitaxial bridging layer that eliminates substrate clamping by providing out-of-plane structural distortions akin to those in the primary structural order parameter. The model material in this work, *h*-$LuFeO_3$, is an improper ferroelectric with a non-polar, trimer structural distortion as its primary order parameter; it is isostructural to *h*-$YMnO_3$ and *h*-$ScFeO_3$. Below the ferroelectric Curie temperature, this structural distortion manifests as "down-down-up" or "up-up-down" rumpling of the lutetium ions in the Lu planes that creates a net polarization along the *c*-axis[19] (Extended Data Fig. 1). The unit cell of the ferroelectric phase is larger and includes three times as many lutetium ions as the unit cell of the paraelectric phase to capture this lattice trimerization. For the bottom electrode, we use $SrCo_2Ru_4O_{11}$, a metallic ruthenate[20] that is reasonably lattice matched to both the ferroelectric phase of *h*-$LuFeO_3$ film[21,22] (–2.4%) and to the underlying $Sr_{1.03}Ga_{10.81}Mg_{0.58}Zr_{0.58}O_{19}$ (SGMZ)[23] substrate (+0.2%) (Supplementary Section 1).

Importantly, we facilitate the structural distortion in ultrathin *h*-$LuFeO_3$ films by depositing it on an engineered surface with an analogous planar rumpling pattern. The stacking and matching of ferrite building blocks leading to this engineered surface are schematically shown in Fig. 1. They are selected to provide structurally compatible bridging layers between the crystal structures involved. There are no commercially available isostructural substrates to *h*-$LuFeO_3$, much less any that are conductive and could serve as bottom electrodes. So, we start with the recently perfected large-diameter substrate SGMZ[23] upon which we deposit the conductor $SrCo_2Ru_4O_{11}$. Both SGMZ and $SrCo_2Ru_4O_{11}$ have crystal structures belonging to the ferrite family; they share the same building blocks including A layer, B layer, S block, and R block (Fig. 1a) following conventional nomenclature[24,25]. Various ferrite structures are made by stacking these building blocks in different sequences, for example, SGMZ consists of alternating S and R blocks (Fig. 1b), whereas $SrCo_2Ru_4O_{11}$ consists of alternating R blocks and B layers[20] (Fig. 1c). The structural transition is made by terminating the SGMZ at the A layer in an S block and beginning the $SrCo_2Ru_4O_{11}$ at the B layer (Fig. 1d). Thus, this bridging B layer not only combines with the A and B layer underneath to complete the S block in SGMZ, but also starts the BRB*R* stacking sequence in the $SrCo_2Ru_4O_{11}$.

Next comes the crucial transition to the *h*-$LuFeO_3$ (Fig. 1e). We leverage what was learned by Akbashev *et al.* from the growth of *h*-$LuFeO_3$ films that were so iron-rich that intergrowths of $Fe_3O_4$ lamella formed within the *h*-$LuFeO_3$ film without destroying the "up-up-down" rumpling of the adjacent lutetium layers[26]. $Fe_3O_4$ also belongs to the ferrite family; indeed, it is comprised of alternating A and B layers (Fig. 1f). What Akbashev *et al.* observed was that between the terminating B layer of $Fe_3O_4$ and the starting $(FeO)^+$ layer of *h*-$LuFeO_3$, a $Lu_{2/3}Fe_{1/3}O_{7/6}$ monolayer naturally formed in their intergrown sample and bridged the two crystal structures (Fig. 1g). This $Lu_{2/3}Fe_{1/3}O_{7/6}$ layer resembles the A layer. Following this knowledge, we synthesize a structurally compatible bridge from $SrCo_2Ru_4O_{11}$ to *h*-$LuFeO_3$ by ending the $SrCo_2Ru_4O_{11}$ at the B layer, adding an A-like $Lu_{2/3}Fe_{1/3}O_{7/6}$ layer, and then starting the 0.75-unit-cell-thick *h*-$LuFeO_3$ film with the $(FeO)^+$ layer (Fig. 1h). The whole stack is grown by reactive-oxide molecular-beam epitaxy (MBE, see Methods), which enables the targeted structure to be built with atomic-layer precision. X-ray diffraction (XRD) and atomic force microscopy (AFM) results (Extended Data Fig. 2a,b,c) show the excellent structural quality and film uniformity that accompanies this logical transition between structures following a growth path consisting of a compatible sequence of layers.

The atomic-level structural schematics and high-angle annular dark field scanning transmission electron microscopy (HAADF-STEM) images of each interface and the entire stack of layers are shown in Fig. 2. Abrupt interfaces are observed between the SGMZ substrate, $SrCo_2Ru_4O_{11}$ bottom electrode, the monolayer-thick $Lu_{2/3}Fe_{1/3}O_{7/6}$ bridging layer, and the 0.75-unit-cell-thick *h*-$LuFeO_3$ film that contains a single $(LuO_2)^-$ monolayer sandwiched between two $(FeO)^+$ layers. Each bridging layer is atomically resolved and follows the designed stacking sequence as labeled in Fig. 2a.

The clear characteristic "down-down-up" planar rumpling pattern of lutetium cations shown in Fig. 2b,d demonstrates that stable improper ferroelectricity persists at the two-dimensional limit at room temperature. The spontaneous rumpling of the lutetium ions that leads to ferroelectricity can be quantified by measuring the amplitude of trimerization, $Q$, in HAADF-STEM images[27]. From the established theory[28], "a non-zero trimer distortion induces a non-zero polarization," and there is a reliable relationship between $Q$ and the spontaneous polarization ($P$) in *h*-$LuFeO_3$. Thus, by measuring $Q$ values, we can map the spontaneous polarization ($P$) of our *h*-$LuFeO_3$ films, as demonstrated in a previous study[29]. The $Q$ is found to be $25 \pm 1$ pm in the 0.75-unit-cell-thick *h*-$LuFeO_3$ film, comparable to the $Q$ in bulk films[30]. This measured $Q$ corresponds to a spontaneous polarization value of $6.4\pm0.3$ μC/cm$^2$ (Supplementary Section 2). Contrary to greatly reduced $Q$ in the first one- or two-unit cells in clamped films[12,13], our 0.75-unit-cell-thick $h$-$LuFeO_3$ film exhibits lattice trimerization and spontaneous polarization retention to more than 80% of the full extent (Supplementary Section 2). The monolayer bridging layer is made up of a repetition of rumpled "Lu-Lu-Fe"[26]. Its cationic rumpling is coherent with that of $h$-$LuFeO_3$, where the lutetium ions in the $h$-$LuFeO_3$ have the same "up" or "down" displacements as the lutetium ions immediately below in the monolayer bridging layer (Fig. 2b,d). This coherency might stabilize and facilitate the ferroelectric structural distortion in our ultrathin $h$-$LuFeO_3$ films. The ferroelectric order in ultrathin $h$-$LuFeO_3$ films were further investigated and corroborated through X-ray linear dichroism (XLD) (Supplementary Section 3).

*In-situ* reflection high-energy electron diffraction (RHEED) is employed to corroborate the onset of the geometric ferroelectricity. The onset of ferroelectricity can be observed *in situ* during growth using RHEED as the ferroelectric structural rumpling of the lutetium ions causes an in-plane tripling of the unit cell ($\sqrt{3} \times \sqrt{3}$ reconstruction in the *a*-*b* plane), which appears in RHEED as streaks at one third of the in-plane spacing of the paraelectric phase[13,31]. Second harmonic generation is used to rule out the possibility that this characteristic RHEED pattern originates from simply a surface effect (Supplementary Section 4). We observe clear one-third-order streaks corresponding to the ferroelectric phase of the 0.75-unit-cell-thick $h$-$LuFeO_3$ film at its growth temperature of 1000 K as shown in the RHEED image (Extended Data Fig. 2d). The growth temperature of 1000 K, measured by an optical pyrometer, is comparable to the ferroelectric Curie temperature of bulk $h$-$LuFeO_3$[22]. This offers further evidence that our engineered template stabilizes improper ferroelectricity at the two-dimensional limit with an undiminished Curie temperature.

Polarization reversal in the ultrathin *h*-$LuFeO_3$ films has been observed by piezoresponse force microscopy (PFM). To account for possible artifacts associated with the electrical charge deposition and injection, a combination of poling by an electrical bias and mechanical force[32] has been used. Fig. 3a,b show PFM images obtained from the 0.75-unit-cell-thick *h*-$LuFeO_3$ free surface where a 2×2 μm$^2$ region is polarized electrically by ±4 V followed by mechanical poling of a 1×1 μm$^2$ center square region with a loading force that was incrementally increased from 150 nN to 1500 nN. In pristine *h*-$LuFeO_3$ film, the PFM signal is relatively weak, while clear PFM contrast is observed in the regions subjected to electrical bias indicating successful poling. Subsequent mechanical switching from the upward to the downward polarization state is manifested as the PFM contrast change in the left half of the 1×1 μm$^2$ center region. PFM switching spectroscopy (Fig. 3c) reveals a conventional butterfly-like strain loop confirming the ferroelectric switching that starts at a coercive voltage of around 2.5 V. Polarization reversal was also observed via PFM imaging of the 2.25, 3.25, and 12.25-unit-cell-thick *h*-$LuFeO_3$ films (Supplementary Section 5). In the 12.25-unit-cell-thick *h*-$LuFeO_3$ film, the polarization reversal was additionally verified by switching current measurements (Supplementary Section 5). Electrical testing in the thinner *h*-$LuFeO_3$ film capacitors is not possible at room temperature due to high leakage.

We attribute the realization of improper ferroelectricity in ultrathin *h*-$LuFeO_3$ films to the similar structural motif compared to *h*-$LuFeO_3$ that is provided by our engineered epitaxial template. In-plane, the primitive unit cell of $SrCo_2Ru_4O_{11}$ is well lattice matched to the ferroelectric phase of *h*-$LuFeO_3$ film (–2.4%). Out-of-plane, the $Lu_{2/3}Fe_{1/3}O_{7/6}$ bridging layer exhibits a cationic planar rumpling pattern that not only shares the same periodicity, but also extends along the identical crystallographic $[1\bar{1}00]$ direction as that of *h*-$LuFeO_3$ (Extended Data Fig. 3). Thus, our epitaxial template possesses a highly analogous structural motif to *h*-$LuFeO_3$ both in-plane and out-of-plane. On the contrary, typically used epitaxial templates for *h*-$LuFeO_3$ and other isostructural improper ferroelectrics, such as (111)-orientated 9.5 mol% yttria-stabilized cubic zirconia (YSZ) substrates[12,21], (111) Ir bottom electrodes[30], (111) Pt bottom electrodes[31,33], etc., offer neither a good structural match in-plane nor out-of-plane. The primitive unit cells of these heretofore utilized substrates are much better lattice matched to the paraelectric phase unit cell than to the ferroelectric phase unit cell of *h*-$LuFeO_3$ and their flat atomic layers lack planar rumpling as shown in Fig. 4. This poor structural compatibility likely underlies the observed substrate clamping and the destruction of the structural distortion in the first one- or two-unit cells near the interface. When we grow *h*-$LuFeO_3$ by MBE on bare or iridium-coated (111) YSZ substrates, we find that RHEED does not show the characteristic ferroelectric feature, i.e., one-third-order streaks, until the *h*-$LuFeO_3$ films are ≥ 1.5-unit cells thick on (111) Ir or 4-unit cells thick on (111) YSZ substrates (Extended Data Fig. 4). Further, STEM studies show a greatly diminished trimerization amplitude in the first few $(LuO_2)^-$ layers in samples on (111) Ir or (111) YSZ substrates (Extended Data Fig. 5). In contrast, in a 2.25-unit-cell-thick *h*-$LuFeO_3$ film grown on the engineered template, the very first $(LuO_2)^-$ layer exhibits trimerization amplitude at its full extent (Extended Data Fig. 5 and Supplementary Section 2).

Ideally, 0.5-unit-cell-thick *h*-$LuFeO_3$ is the thinnest form of this material, containing only one layer of iron oxide and one layer of lutetium oxide. Attempts to further reduce the film thickness to 0.5-unit cells fail because the lutetium plane is found to be unstable for terminating *h*-$LuFeO_3$[34] (see HAADF-STEM evidence in Extended Data Fig. 6). Attempts to deposit *h*-$LuFeO_3$ directly on the B layer of $SrCo_2Ru_4O_{11}$, i.e., without the $Lu_{2/3}Fe_{1/3}O_{7/6}$ bridging layer, result in the observation of the $Lu_{2/3}Fe_{1/3}O_{7/6}$ monolayer in the grown film by HAADF-STEM (Extended Data Fig. 7). The observed $Lu_{2/3}Fe_{1/3}O_{7/6}$ monolayer forms spontaneously and because those atoms diffuse out of the overlying *h*-$LuFeO_3$ layers, the overlying *h*-$LuFeO_3$ film is lutetium deficient, resulting in the defective microstructure seen. This spontaneous formation of the $Lu_{2/3}Fe_{1/3}O_{7/6}$ monolayer demonstrates its natural bridging function and shows the robustness of our strategy to bridge distinct oxide structures[35].

Having established a robust route to achieve ferroelectricity at the monolayer limit of *h*-$LuFeO_3$, we apply this approach to growth on conventional *c*-plane sapphire substrates. Extended Data Figs. 8 and 9 show XRD, RHEED, AFM, and HAADF-STEM results demonstrating ferroelectricity in a 0.75-unit-cell-thick *h*-$LuFeO_3$ film grown using a $Lu_{2/3}Fe_{1/3}O_{7/6}$ monolayer bridging layer on a $SrCo_2Ru_4O_{11}$ bottom electrode on a (0001) sapphire substrate.

Overall, we have engineered the thinnest improper ferroelectric film, 0.75-unit-cell-thick *h*-$LuFeO_3$, with undiminished polarization and Curie temperature compared to bulk films. In particular, we find that the cationic rumpling pattern in a monolayer bridging layer can stabilize and facilitate the ferroelectric structural distortion in the thinnest possible *h*-$LuFeO_3$ films. Importantly, this robust ultrathin ferroelectric can also be made on widely available sapphire substrates. This work not only demonstrates the absence of a critical thickness for improper ferroelectricity, but also presents a methodology for creating ultrathin improper ferroelectric films by stabilizing the primary order parameter. This understanding could guide the design of more ultrathin improper ferroelectrics for device miniaturization. More broadly, our results point to the importance of considering structural compatibility beyond lattice match when transitioning between different structure types.

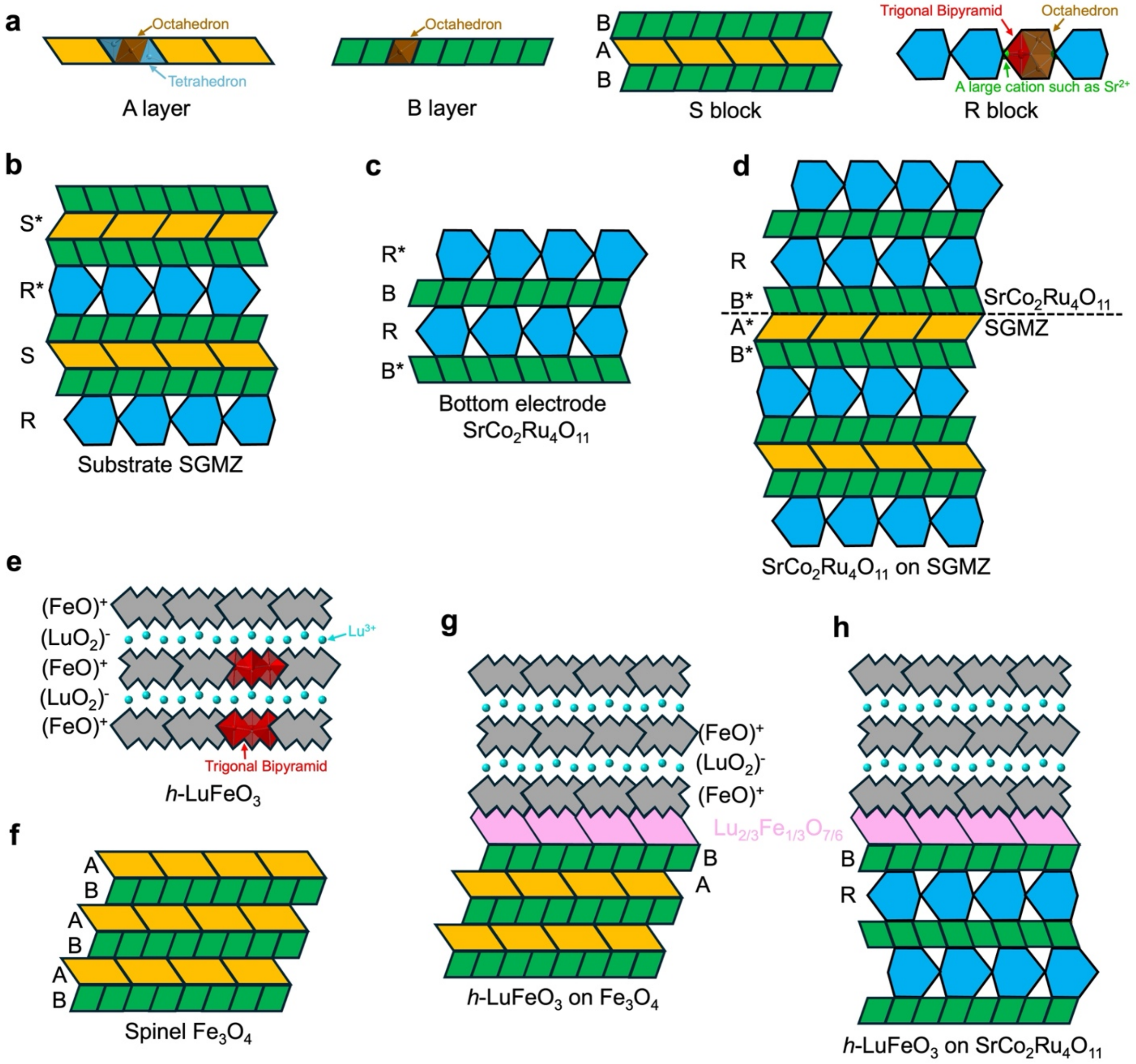

**Figure 1 | Schematic crystal structure building and interfacing. a,** Basic building blocks for ferrites. **b,** Alternating S and R blocks in the crystal structure of the SGMZ substrate with the [0001] direction vertical and $[11\bar{2}0]$ into the paper. **c,** Alternating B layers and R blocks in the crystal structure of the bottom electrode $SrCo_2Ru_4O_{11}$ with the [0001] direction vertical and $[11\bar{2}0]$ into

the paper. **d,** Structural interfacing between $SrCo_2Ru_4O_{11}$ and SGMZ. **e,** Alternating $(FeO)^+$ layers and $(LuO_2)^-$ layers in the crystal structure of $h$-$LuFeO_3$ with the [0001] direction vertical and $[11\bar{2}0]$ into the paper. **f,** Alternating A and B layers in the crystal structure of spinel $Fe_3O_4$ with the $[\bar{1}\bar{1}\bar{1}]$ direction vertical and $[\bar{1}01]$ into the paper. **g,** Structural interfacing between $h$-$LuFeO_3$ and $Fe_3O_4$ via a $Lu_{2/3}Fe_{1/3}O_{7/6}$ bridging layer. **h,** Structural interfacing between $h$-$LuFeO_3$ and $SrCo_2Ru_4O_{11}$ via a $Lu_{2/3}Fe_{1/3}O_{7/6}$ bridging layer. (The star symbol * means that the block/layer is rotated 180° around the *c* axis with respect to the block/layer without the star.)

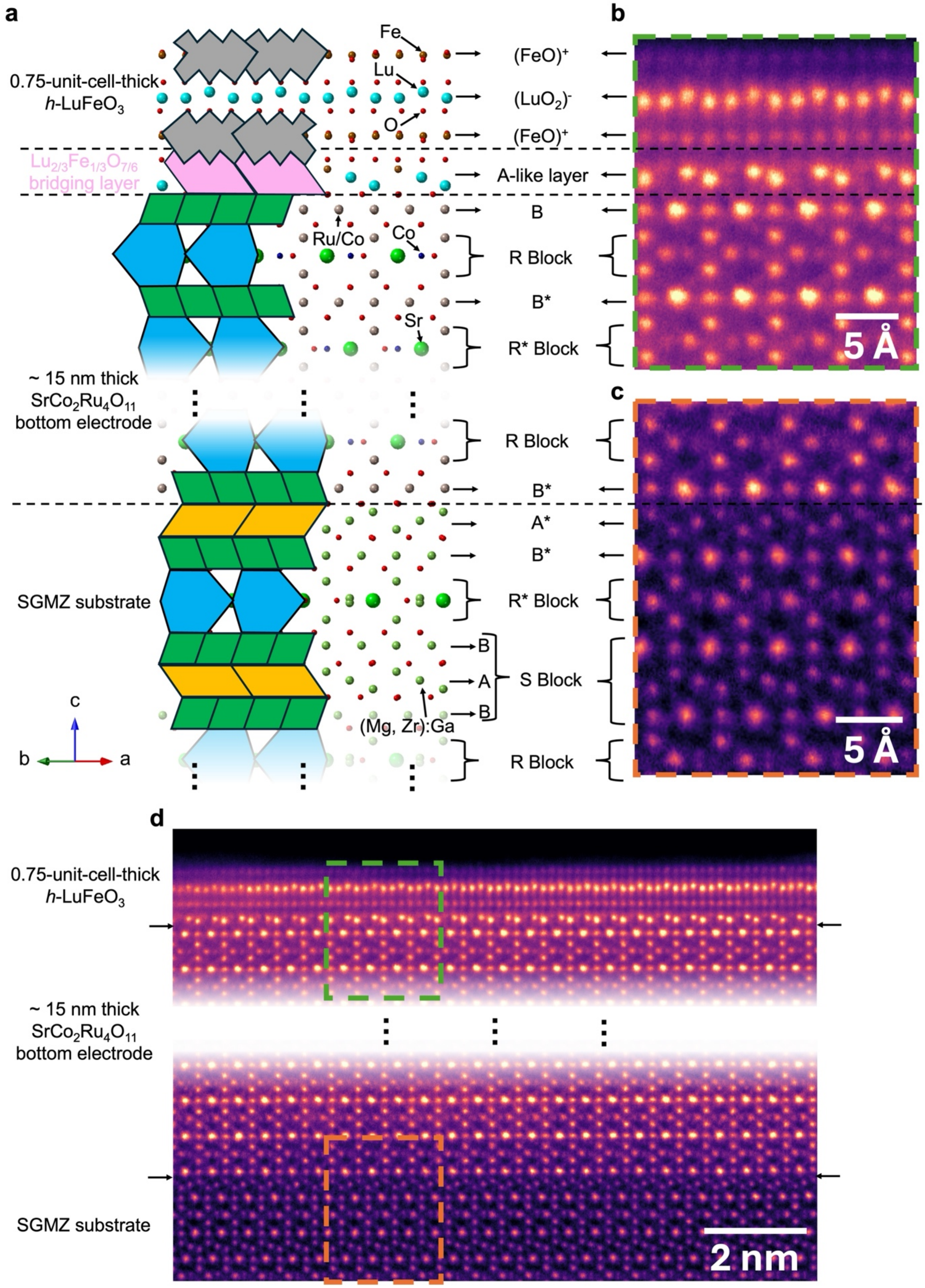

**Figure 2 | Schematic crystal structure and HAADF-STEM images showing atomic-level details of a 0.75-unit-cell-thick *h*-$LuFeO_3$ film on the engineered epitaxial template. a,** Schematic atomic structure of the entire stack, 0.75-unit-cell-thick *h*-$LuFeO_3$/$Lu_{2/3}Fe_{1/3}O_{7/6}$ monolayer/$SrCo_2Ru_4O_{11}$ bottom electrode/SGMZ substrate. (The star symbol * means that the block/layer is rotated 180° around the *c* axis with respect to the block/layer without the star.) **b-d,** HAADF-STEM images of **b,** the structural transition from $SrCo_2Ru_4O_{11}$ to *h*-$LuFeO_3$, **c,** the structural transition from the SGMZ substrate to $SrCo_2Ru_4O_{11}$, and **d,** the entire stack. **b** and **c** are magnified from the green boxed region and the orange boxed region in **d**, respectively.

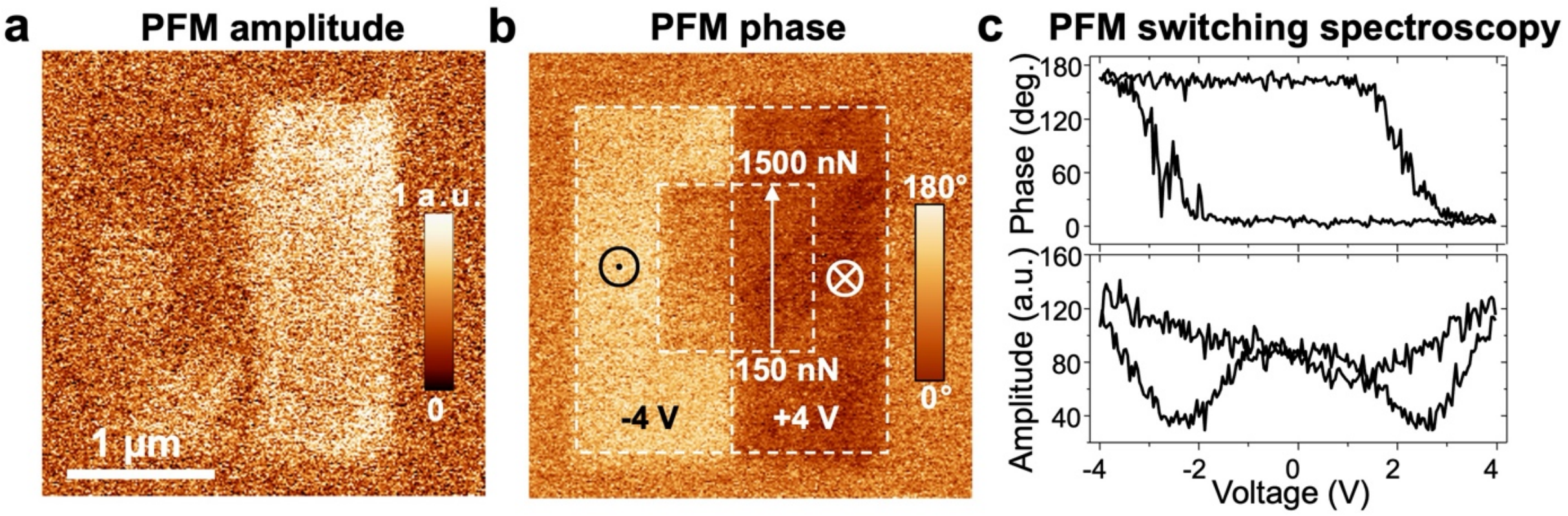

**Figure 3 | Observation of ferroelectric switching in the 0.75-unit-cell-thick *h*-$LuFeO_3$ film via PFM measurements. a-b**, PFM amplitude and phase images obtained after electrical poling by ±4 V in a 2×2 μ$m^2$ region (left side, -4 V, and right side, +4 V), followed by mechanical poling in a 1×1 μ$m^2$ center square region using a mechanical loading force varied in the range from 150 nN to 1500 nN. **c**, Representative PFM hysteresis loops obtained at a fixed location in the *h*-$LuFeO_3$ film.

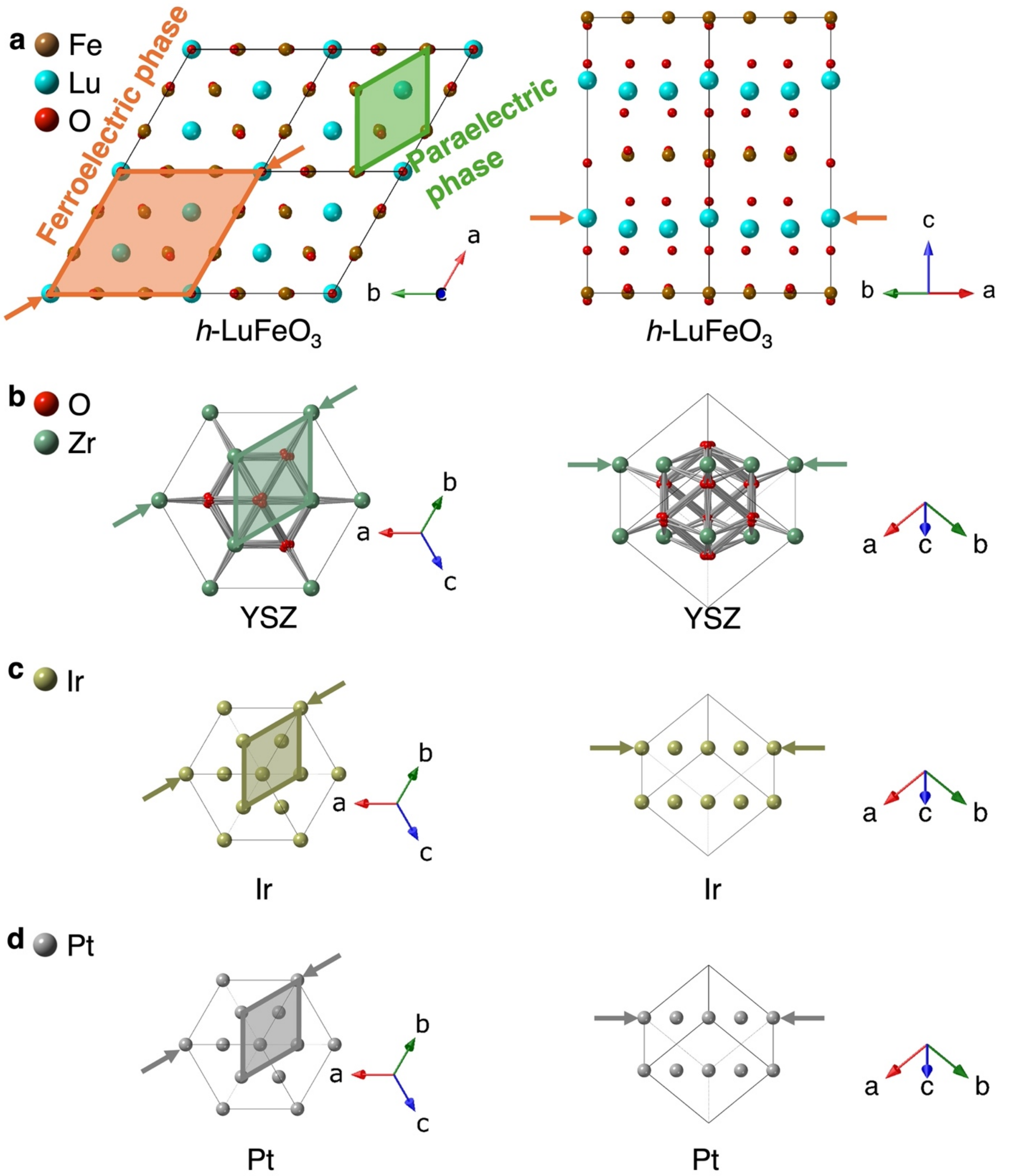

**Figure 4 | In-plane and out-of-plane structure motif comparison between *h*-LuFeO₃ and some typically used epitaxial templates for *h*-LuFeO₃ and other isostructural improper ferroelectrics. a.** Ferroelectric phase (orange) and paraelectric phase (green) unit cells of *h*-

$LuFeO_3$, projected onto the *a*-*b* plane (left). Crystal structure of $h$-$LuFeO_3$ viewed along $[11\bar{2}0]$ (right). **b,** YSZ primitive unit cell, projected onto the (111) plane (left). Crystal structure of YSZ viewed along $[11\bar{2}]$ (right). **c,** Ir primitive unit cell, projected onto the (111) plane (left). Crystal structure of Ir viewed along $[11\bar{2}]$ (right). **d,** Pt primitive unit cell, projected onto the (111) plane (left). Crystal structure of Pt viewed along $[11\bar{2}]$ (right). The sizes of the unit cells are to scale based on their lattice parameters.

# Methods

**Growth of $SrCo_2Ru_4O_{11}$ bottom electrodes.** Thin films of $SrCo_2Ru_4O_{11}$ bottom electrodes were grown by reactive-oxide molecular-beam epitaxy (MBE) in a Veeco GEN10 system on (0001) hexagallate substrates $Sr_{1.03}Ga_{10.81}Mg_{0.58}Zr_{0.58}O_{19}$ (SGMZ) and (0001) sapphire substrates. Strontium and cobalt were evaporated from elemental sources at fluxes of $2 \times 10^{12}$ and $4 \times 10^{12}$ atoms $cm^{-2}$ $s^{-1}$, respectively. Flux calibration methods for strontium and cobalt with an absolute accuracy of ±1% were described in our previous work[36]. A molecular beam of ruthenium was generated from an electron-beam evaporator and its flux was checked by a quartz crystal microbalance (QCM). In an adsorption-controlled growth regime[37], excess ruthenium can desorb from the growth front in the form of volatile $RuO_x$ ($x$ = 2 or 3) and leave behind phase-pure $SrCo_2Ru_4O_{11}$. Thus, the ruthenium flux we used ranged from $1.0 \times 10^{13}$ to $1.2 \times 10^{13}$ atoms cm$^{-2}$ s$^{-1}$, which exceeded the amount required for a stoichiometric film by 1.25 to 1.5 times. 15 nm thick $SrCo_2Ru_4O_{11}$ films were grown by co-deposition of strontium, cobalt, and ruthenium at a substrate temperature of 500 °C measured by an optical pyrometer operating at a wavelength of 1550 nm. Distilled ozone (~80% $O_3$ + 20% $O_2$) was used as the oxidant and the background pressure of this oxidant was $2.2 \times 10^{-7}$ Torr during growth. *In situ* reflection high-energy electron diffraction (RHEED) patterns were recorded using KSA-400 software and a Staib electron source operated at 13 kV and a filament current of 1.5 A. X-ray diffraction (XRD) scans were measured with a PANalytical Empyrean diffractometer with Cu $K\alpha_1$ radiation. Atomic force microscopy (AFM) surface topography images were taken with the Asylum Research Cypher ES Environmental AFM. After characterization, the bottom electrodes were loaded back to our Veeco GEN10 MBE system for the growth of $Lu_{2/3}Fe_{1/3}O_{7/6}$ monolayer bridging layer and ultrathin *h*-$LuFeO_3$.

**Growth of $Lu_{2/3}Fe_{1/3}O_{7/6}$ monolayer bridging layer and ultrathin *h*-$LuFeO_3$.** Iron and lutetium were evaporated from elemental sources with the same flux of $2 \times 10^{13}$ atoms cm$^{-2}$ s$^{-1}$. Flux calibration methods for iron and lutetium with ±1% accuracy were also described in our previous work[36]. During the growth of the monolayer bridging layer and *h*-$LuFeO_3$, the substrate temperature was kept at ~730 °C measured by an optical pyrometer operating at a wavelength of 1550 nm. And distilled ozone (~80% $O_3$ + 20% $O_2$) was supplied continuously at a background pressure of $5\times10^{-7}$ Torr. The growth of $Lu_{2/3}Fe_{1/3}O_{7/6}$ monolayer bridging layer was initiated by co-depositing one-third monolayer of iron oxide and one-third monolayer of lutetium oxide. Then the iron shutter was closed, and additional one-third monolayer of lutetium oxide was deposited. Alternating monolayers of iron oxide and lutetium oxide were deposited to complete *h*-$LuFeO_3$ samples with desired thicknesses. *In situ* RHEED patterns were recorded using KSA-400 software and a Staib electron source operated at 13 kV and a filament current of 1.5 A. AFM surface topography images were taken with the Asylum Research Cypher ES Environmental AFM.

**Scanning transmission electron microscopy (STEM).** Samples for cross-sectional imaging were prepared using the standard lift-out procedure in a Thermo Fisher Helios G4 UX focused ion beam. HAADF-STEM images were acquired on an aberration-corrected Thermo Fisher Spectra 300 X-CFEG STEM operated at 300 kV with a probe semi-convergence angle of 30 mrad, probe current of 60 pA, and a collection angle of 60-200 mrad. A series of 10-20 images, each with a dwell time of 0.5-1 μs were acquired, registered, and averaged for drift correction.

**Piezoresponse force microscopy (PFM).** PFM measurements were carried out using an Asylum Research MFP-3D atomic force microscopy system. Resonance-enhanced dual frequency tracking PFM mode was used to amplify the PFM signal, where an AC modulation voltage of 0.8 V in amplitude and 365 kHz frequency was applied to the Pt-coated Si tip (PPP-EFM). PFM switching spectroscopy was performed at a fixed location using a triangular waveform with voltage sweeping at 0.25 Hz in the pulsed mode. Mechanical poling was performed by scanning the film surface with the electrically grounded tip under loading forces much higher than that used for PFM measurements (typically in the range of 50 nN~100 nN).

# Data Availability Statement

Additional data related to the growth and structural characterization of the thin film heterostructures is available at https://data.paradim.org/doi/ehb7-nm73

# Acknowledgements

Y.E.L., H.K.P., R.R., D.A.M. and D.G.S. acknowledge support from the Army Research Office under the ETHOS MURI via cooperative agreement W911NF-21-2-0162. M.B., C.G., and D.G.S. acknowledge the Joint Lab between Cornell University and the Leibniz-Institut für Kristallzüchtung that facilitated this research. The electron microscopy studies made use of the Cornell Center for Materials Research (CCMR) facilities supported by NSF MRSEC program (DMR-1719875), NSF MIP (DMR-2039380), NSF-MRI-1429155 and NSF (DMR-1539918). The authors also thank John Grazul, Mariena Silvestry Ramos and Malcolm Thomas for technical support and maintenance of the electron microscopy facilities. A.G. acknowledges support by the UNL Grand Challenges catalyst award "Quantum Approaches Addressing Global Threats." S. H. and V. G. acknowledge support from the Department of Energy, Basic Energy Sciences grant number DE-SC0012375 for the optical second harmonic generation measurements. A.L. acknowledges support from National Science Foundation (REU Site: Summer Research Program at PARADIM) under cooperative agreement DMR-2150446.

# Competing Interests Declaration

The authors R.A.S., M.E.H. and D.G.S. have been granted U.S. Patent No. 11,678,587 B2 (13 June 2023) with the title "Ferroelectric and Multiferroic Material Structures."

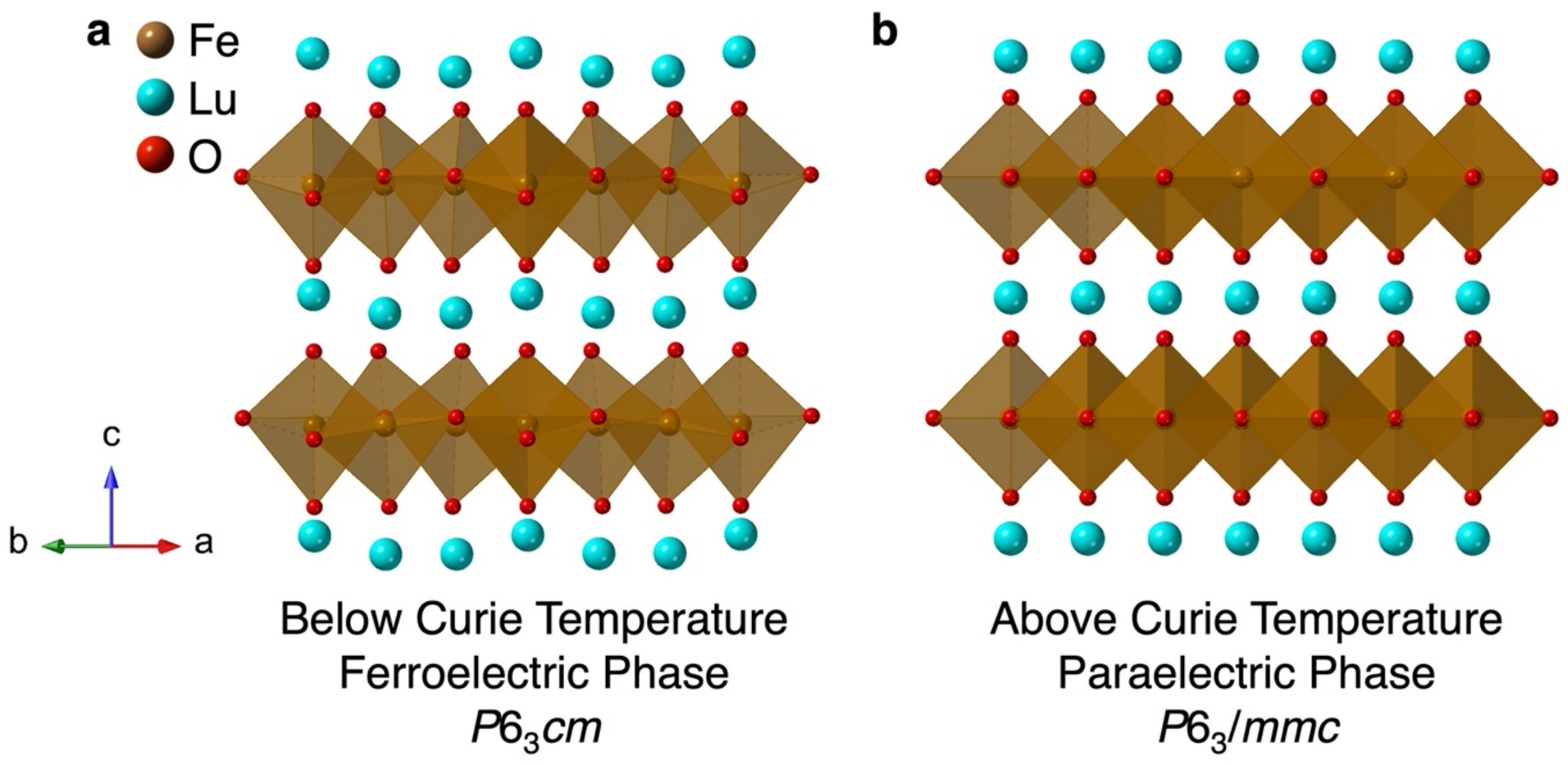

**Extended Data Figure 1 | Schematic atomic structures of *h*-$LuFeO_3$. a,** The ferroelectric phase. **b,** The paraelectric phase.

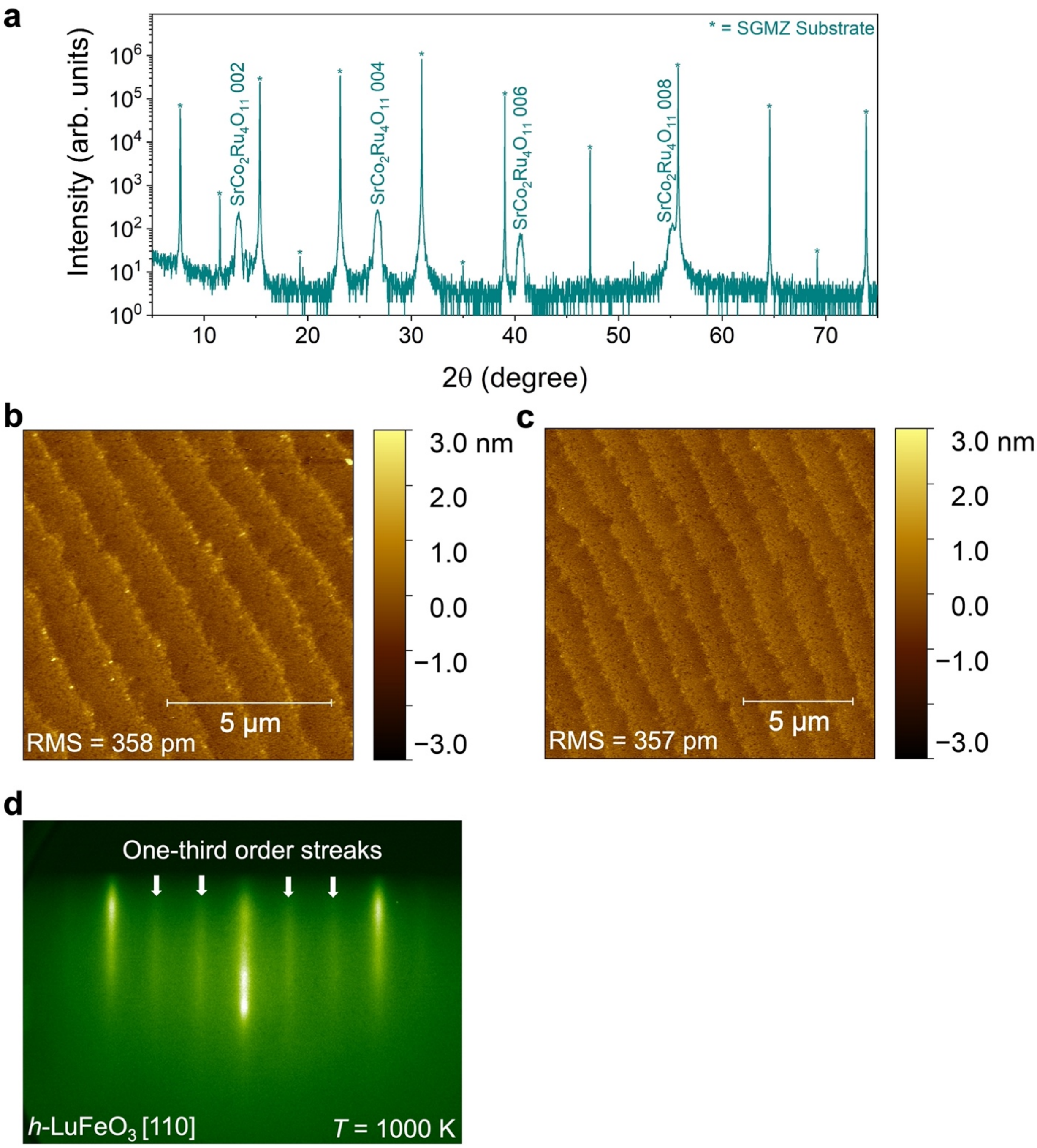

**Extended Data Figure 2 | Characterizations of a 0.75-unit-cell-thick *h*-$LuFeO_3$ film on the engineered epitaxial template. a,** $\theta$–$2\theta$ XRD scan of the $SrCo_2Ru_4O_{11}$ bottom electrode. The asterisk (*) indicates the 000*n* peaks from the (0001) SGMZ substrate. **b,** AFM scan of the $SrCo_2Ru_4O_{11}$ bottom electrode before the growth of the $Lu_{2/3}Fe_{1/3}O_{7/6}$ monolayer and the 0.75-

unit-cell-thick *h*-$LuFeO_3$ film. **c,** AFM scan after the growth of the $Lu_{2/3}Fe_{1/3}O_{7/6}$ monolayer and the 0.75-unit-cell thick *h*-$LuFeO_3$ film. AFM images of the bare $SrCo_2Ru_4O_{11}$ bottom electrode and the *h*-$LuFeO_3$ film all show clear atomic terraces and equally smooth surfaces, suggesting the *h*-$LuFeO_3$ film completely and uniformly covers the bottom electrode. **d,** One-third-order streaks in the *in-situ* RHEED image at the end of growth showing the structural in-plane tripling of the unit cell leading to ferroelectricity. This image was taken at 1000 K, measured by an optical pyrometer.

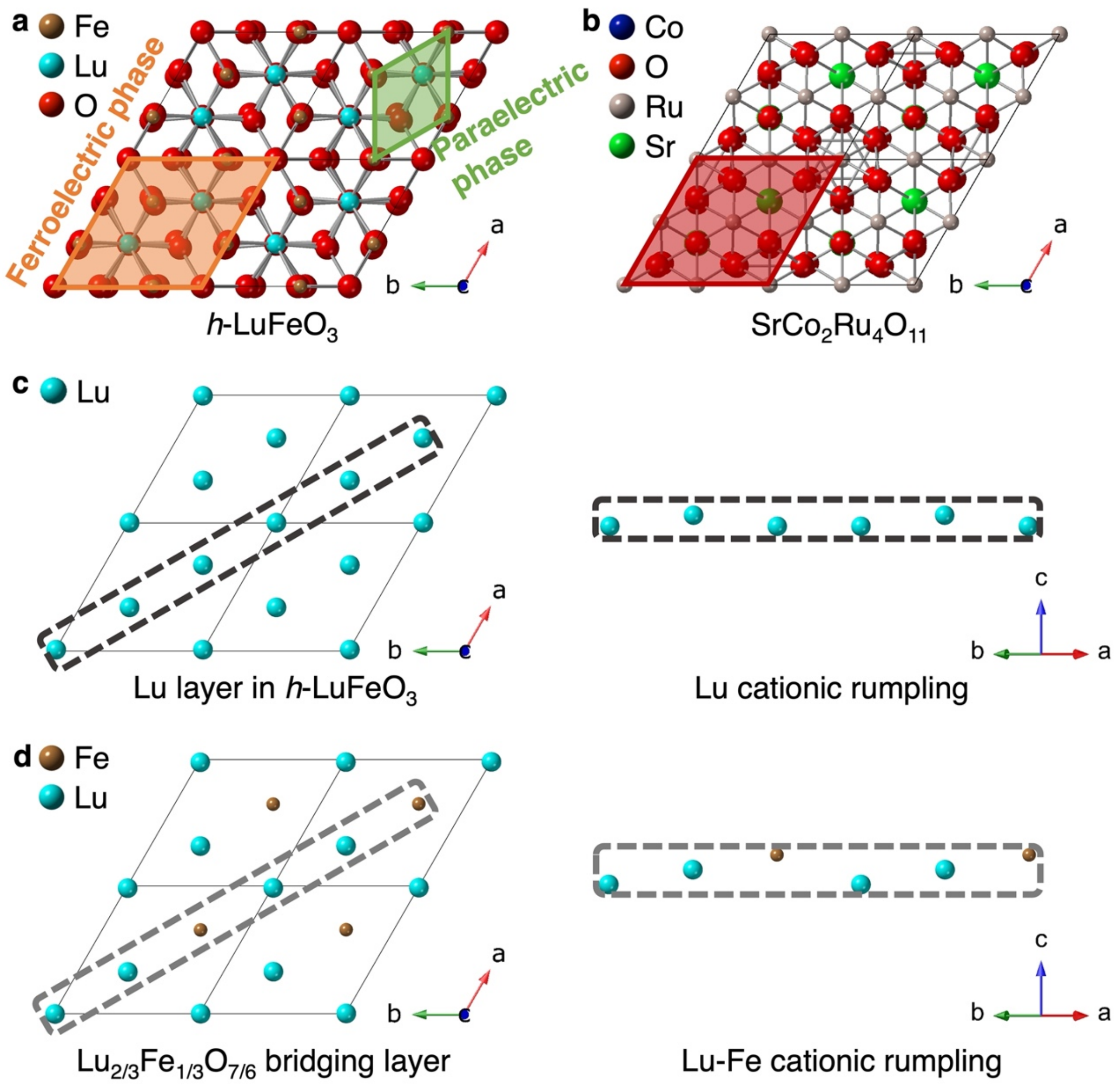

**Extended Data Figure 3 | Unit cell schematics showing that our epitaxial template provides a highly analogous structural motif to *h*-$LuFeO_3$ both in-plane and out-of-plane. a,** Ferroelectric phase (orange) and paraelectric phase (green) unit cells of *h*-$LuFeO_3$, projected onto the *a*-*b* plane. **b,** $SrCo_2Ru_4O_{11}$ unit cell, projected onto the *a*-*b* plane. **c,** *a*-*b* plane projected Lu layer of ferroelectric *h*-$LuFeO_3$ and its cationic rumpling viewed along [110] (Oxygen atoms are not plotted). **d,** *a*-*b* plane projected $Lu_{2/3}Fe_{1/3}O_{7/6}$ bridging layer and its cationic rumpling viewed along [110] (Oxygen atoms are not plotted). The sizes of the unit cells are to scale based on their lattice parameters.

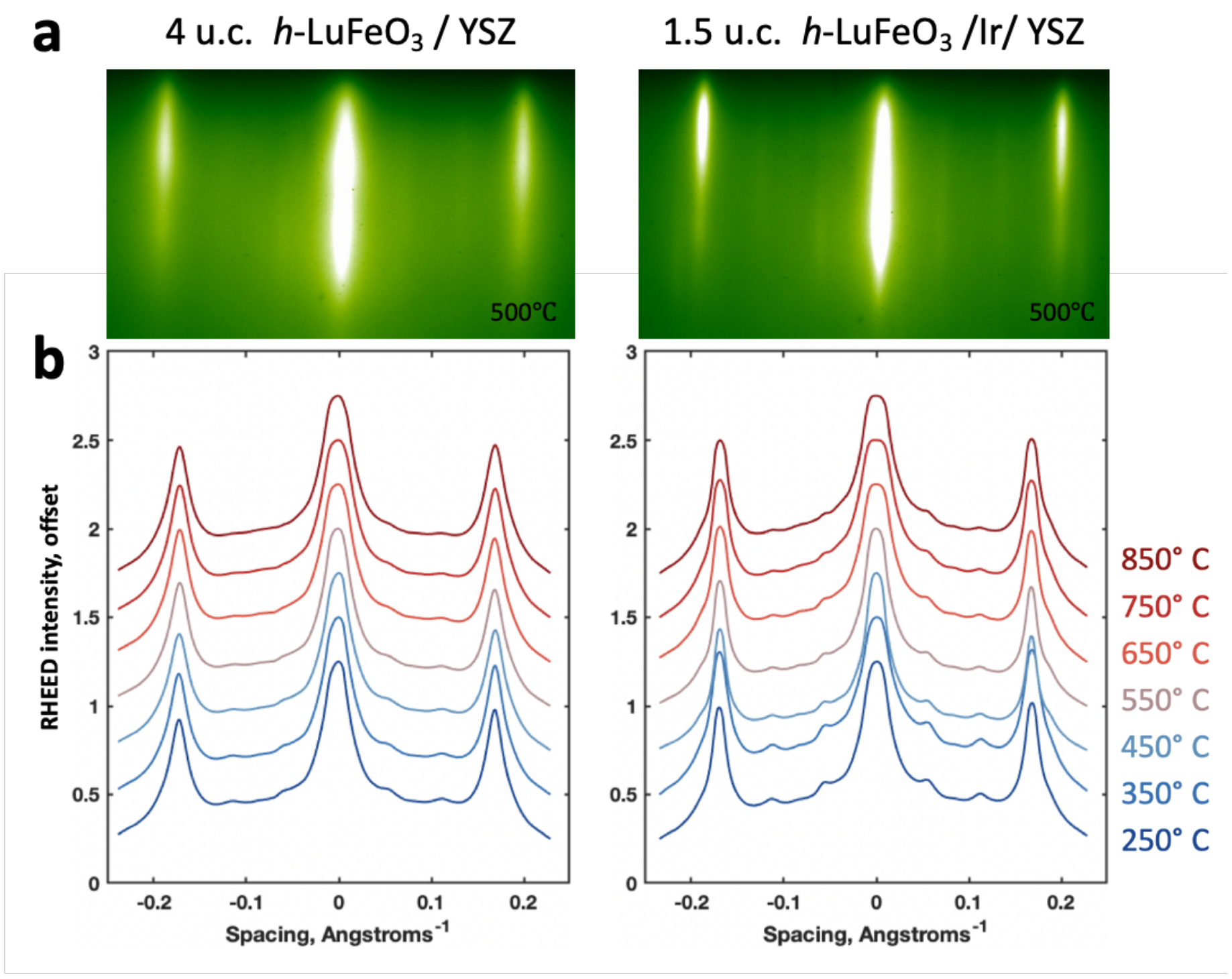

**Extended Data Figure 4 | *In-situ* RHEED images and line scans showing the structural in-plane tripling of the unit cell leading to ferroelectricity.** **a**, RHEED images of a 4-unit-cell-thick *h*-$LuFeO_3$ film grown directly on YSZ (left) and a 1.5-unit-cell-thick *h*-$LuFeO_3$ film grown on an iridium electrode (right). Both images were taken at 500 °C. **b**, Line scans of the RHEED intensity across the image for these films during cooling. The temperatures indicated are the thermocouple temperatures; the true substrate temperature is lower (about 100 °C lower at the highest temperature indicated).

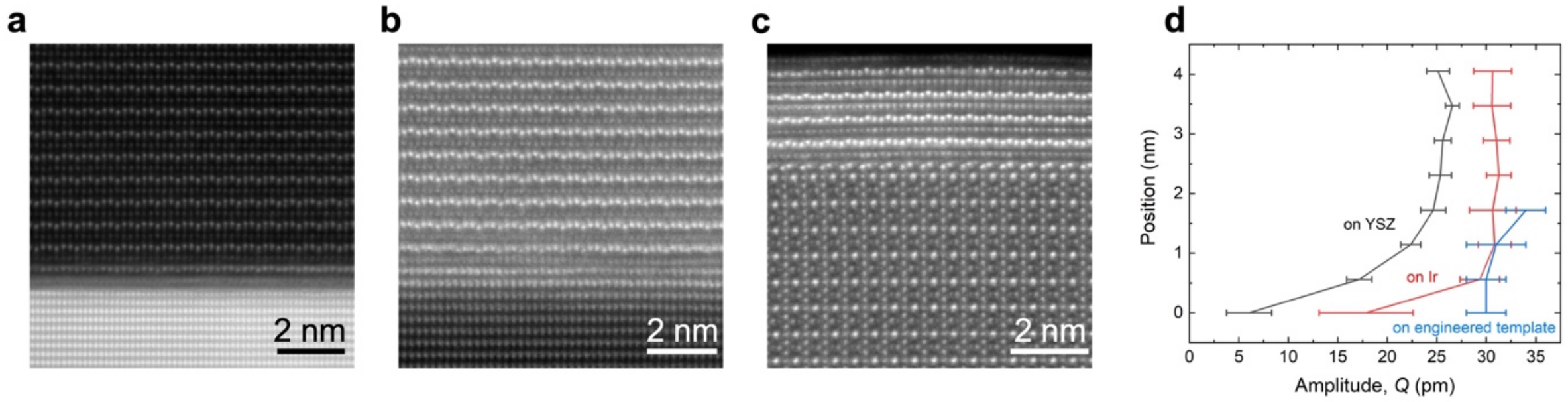

**Extended Data Figure 5 | HAADF-STEM data comparing the interfaces of *h*-$LuFeO_3$ grown on an iridium electrode, directly on a (111) YSZ substrate, and on the engineered template**. **a**, The interface of a *h*-$LuFeO_3$ film grown on an iridium electrode. **b**, The interface of a *h*-$LuFeO_3$ film grown directly on the (111) YSZ substrate. **c**, The interface of a 2.25-unit-cell-thick *h*-$LuFeO_3$ film grown on the engineered template. **d**, $Q$ amplitude of the lutetium-ion distortion calculated monolayer-by-monolayer for *h*-$LuFeO_3$ grown on Ir coated (111) YSZ (red), directly on (111) YSZ (black), and on the engineered template (blue). Error bars correspond to the standard deviation.

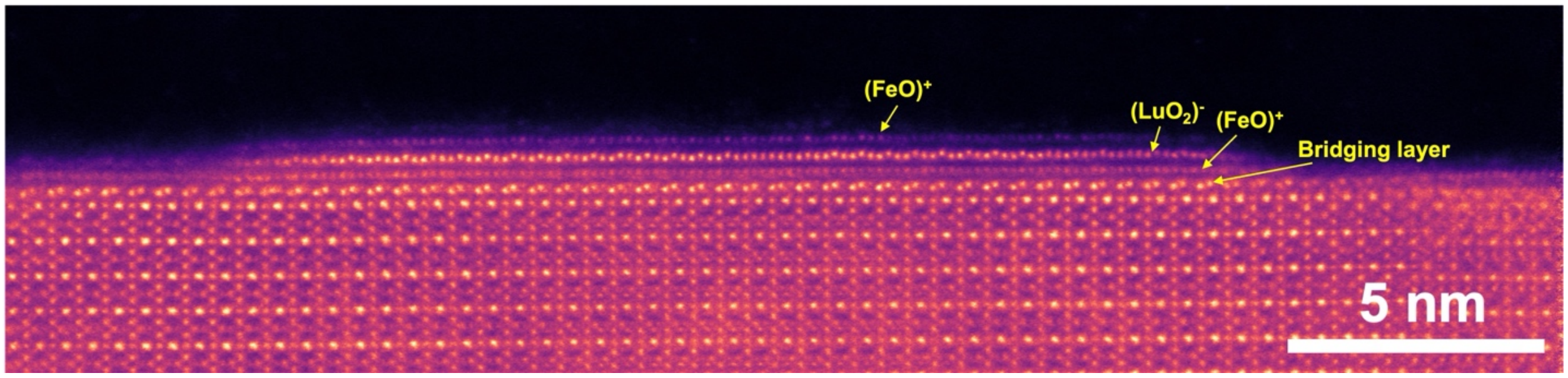

**Extended Data Figure 6 | HAADF-STEM of a *h*-$LuFeO_3$ film that was attempted to be made into a 0.5-unit-cell-thick film.** To make this "0.5-unit-cell-thick" *h*-$LuFeO_3$ film, we deposit one monolayer of iron oxide and one monolayer of lutetium oxide following the $Lu_{2/3}Fe_{1/3}O_{7/6}$ bridging layer. With HAADF-STEM, however, we observe small patches of 0.75-unit-cell-thick *h*-$LuFeO_3$ ending with the iron oxide layer instead of a continuous 0.5-unit-cell-thick *h*-$LuFeO_3$ film terminating with the lutetium oxide layer. So, 0.75-unit cell is the smallest possible thickness of our *h*-$LuFeO_3$ films.

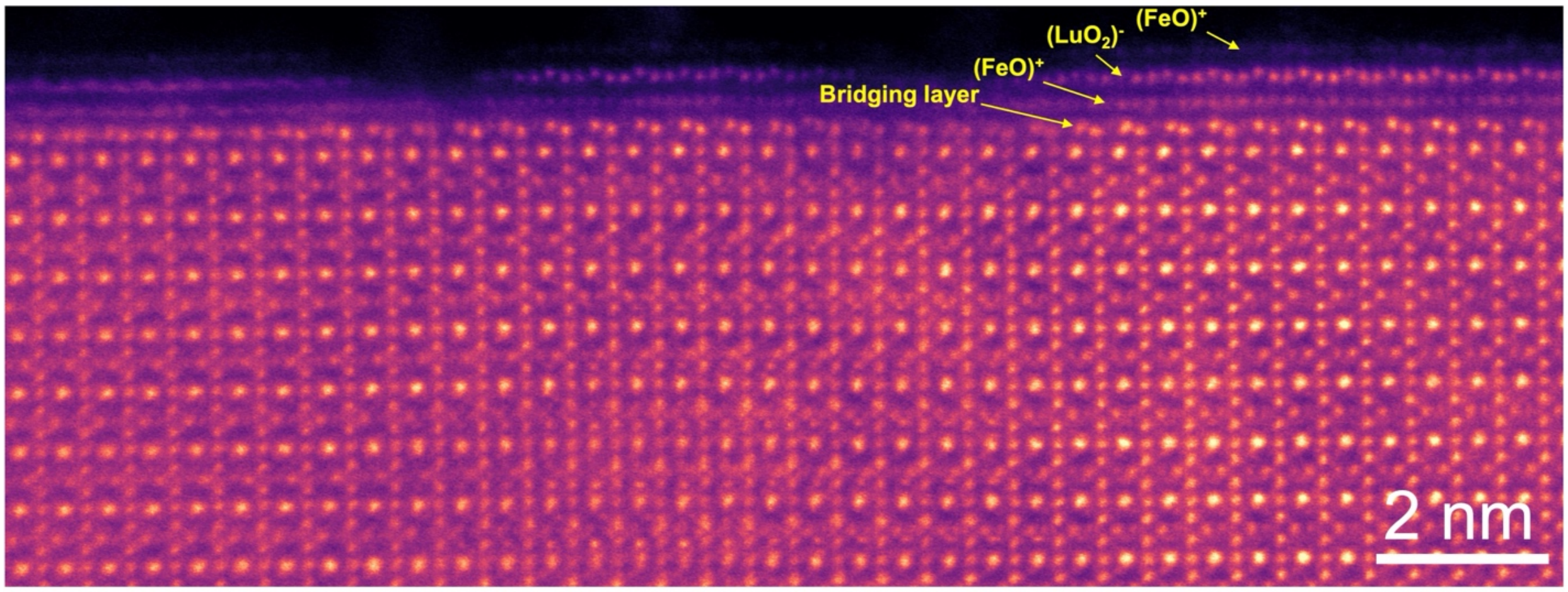

**Extended Data Figure 7 | HAADF-STEM of a "0.75-unit-cell-thick" *h*-$LuFeO_3$ film without intentionally depositing the $Lu_{2/3}Fe_{1/3}O_{7/6}$ bridging layer on a bare $SrCo_2Ru_4O_{11}$ bottom electrode.** On a bare $SrCo_2Ru_4O_{11}$ bottom electrode grown on a SGMZ substrate, we skip the growth of $Lu_{2/3}Fe_{1/3}O_{7/6}$ monolayer bridging layer and directly deposit a 0.75-unit-cell-thick *h*-$LuFeO_3$ film layer-by-layer, starting from the iron oxide layer. Interestingly, the HAADF-STEM image of this sample still shows a clear and continuous $Lu_{2/3}Fe_{1/3}O_{7/6}$ monolayer bridging layer between $SrCo_2Ru_4O_{11}$ bottom electrode and discontinuous *h*-$LuFeO_3$ patches. The *h*-$LuFeO_3$ in this sample is not continuous because some of the lutetium and iron ions provided migrate to the $Lu_{2/3}Fe_{1/3}O_{7/6}$ monolayer bridging layer, leading to a shortage of lutetium and iron ions to form a continuous 0.75-unit-cell-thick *h*-$LuFeO_3$ film.

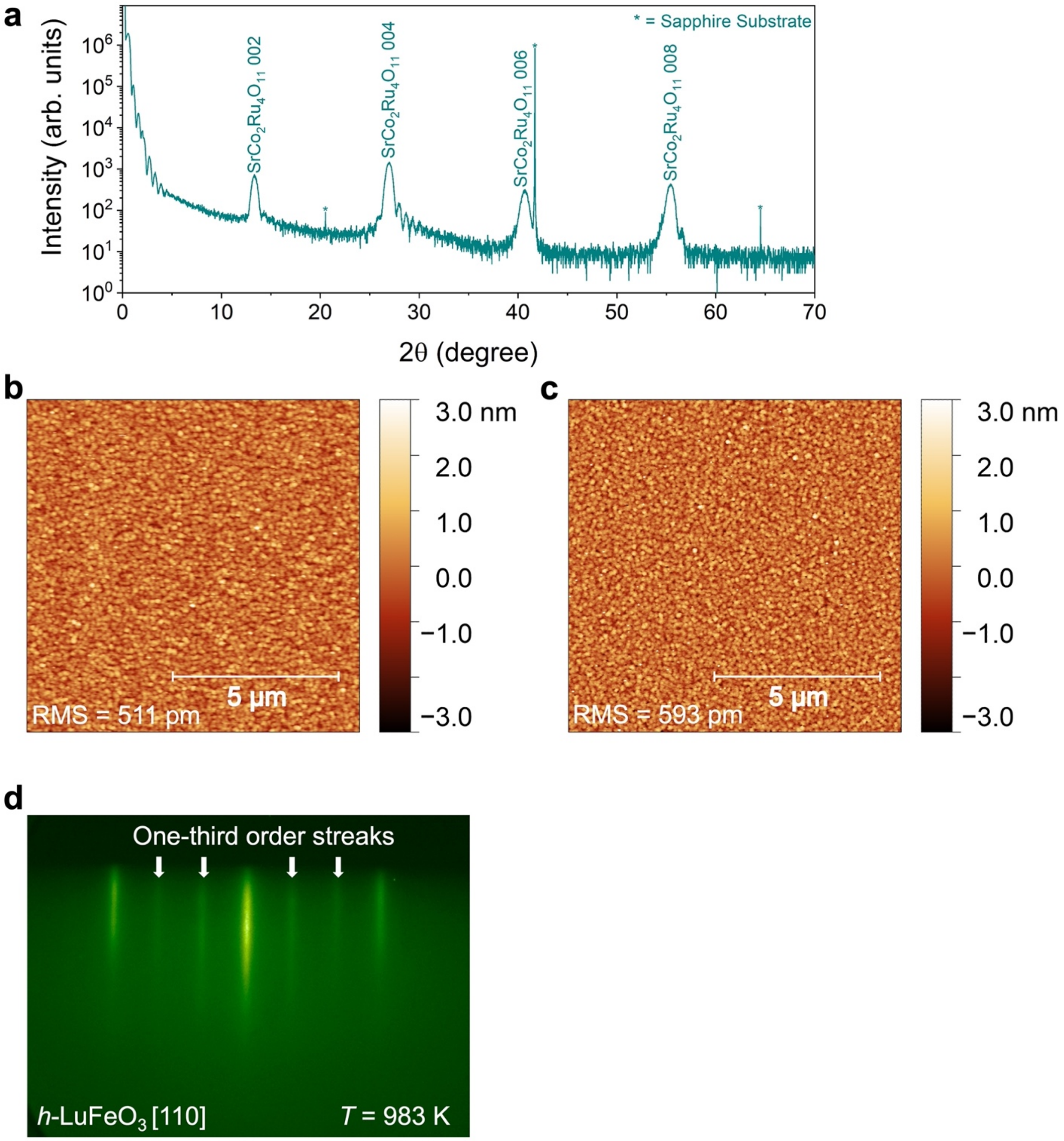

**Extended Data Figure 8 | XRD, AFM, and *in-situ* RHEED characterizations of a 0.75-unit-cell-thick *h*-$LuFeO_3$ using (0001) sapphire as the substrate. a,** $\theta$–$2\theta$ XRD scan of the $SrCo_2Ru_4O_{11}$ bottom electrode. The asterisk (*) indicates the 000*n* peaks from the (0001) sapphire substrate. **b,** AFM scan of the $SrCo_2Ru_4O_{11}$ bottom electrode before the growth of the

$Lu_{2/3}Fe_{1/3}O_{7/6}$ monolayer and the 0.75-unit-cell-thick *h*-$LuFeO_3$ film. **c,** AFM scan after the growth of the $Lu_{2/3}Fe_{1/3}O_{7/6}$ monolayer and the 0.75-unit-cell-thick *h*-$LuFeO_3$ film. AFM images of the bare $SrCo_2Ru_4O_{11}$ bottom electrode and the *h*-$LuFeO_3$ film show comparably smooth surfaces, suggesting the *h*-$LuFeO_3$ film completely and uniformly covers the bottom electrode. **d**, One-third-order streaks in the *in-situ* RHEED image at the end of the growth showing the structural in-plane tripling of the unit cell leading to ferroelectricity. This image was taken at 983 K, measured by an optical pyrometer.

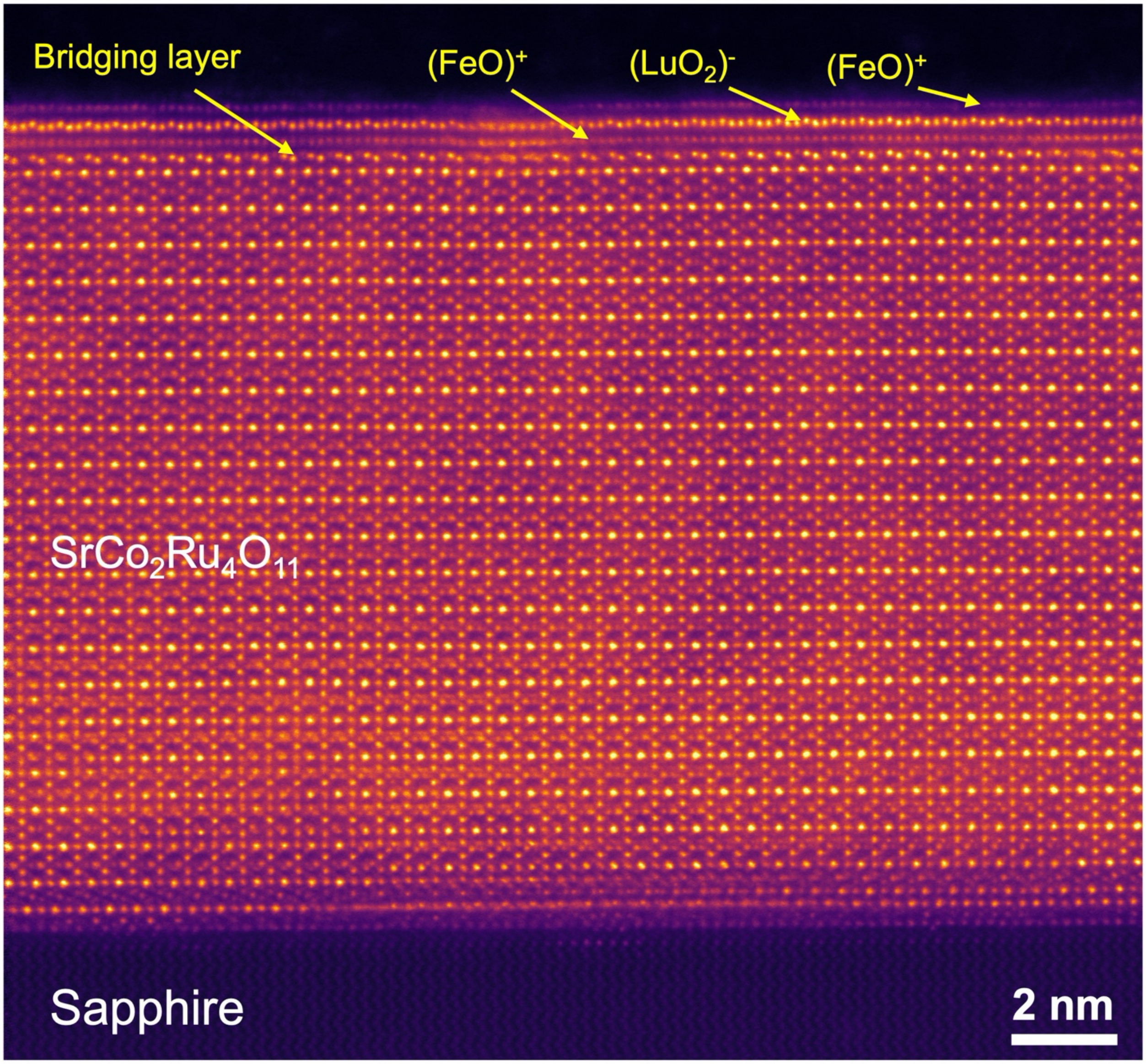

**Extended Data Figure 9 | HAADF-STEM image of the entire stack of a 0.75-unit-cell-thick *h*-$LuFeO_3$ sample on a sapphire substrate.** Abrupt interfaces are observed between the 0.75-unit-cell-thick *h*-$LuFeO_3$ film, the $SrCo_2Ru_4O_{11}$ bottom electrode, and the $Lu_{2/3}Fe_{1/3}O_{7/6}$ monolayer bridging layer. The 0.75-unit-cell-thick *h*-$LuFeO_3$ film is continuous and the "down-down-up" planar rumpling pattern is clear in the lutetium oxide layer.

# Supplementary Information

## Section 1 — Lattice constants of $SrCo_2Ru_4O_{11}$, SGMZ, and *h*-$LuFeO_3$ used for lattice mismatch calculations.

The in-plane lattice constants of single-crystalline $SrCo_2Ru_4O_{11}$ and SGMZ are determined to be about 5.837 Å (Ref. 1) and 5.824 Å (Ref. 2), respectively. Structurally relaxed thick films of *h*-$LuFeO_3$ with the narrowest XRD omega-rocking curves and absence of impurity phases in both structural (AFM and TEM) and magnetic measurements (SQUID) indicate that the unstrained lattice constants of *h*-$LuFeO_3$ are about $a$ = 5.979(5) Å and $c$ = 11.81(3) Å (Ref. 3,4).

## Section 2 — Quantification of the lattice trimerization and the method of mapping electric polarization

Our study specifically measures (by HAADF-STEM) both the lattice trimerization $Q$ and the Lu atomic displacement $d$ as proxies, which directly correlate with the electric polarization $P$ based on theoretical models. Both $Q$ and $d$ are quantitative descriptors of the extent of lattice trimerization and $d = 1.5\ Q$. The schematic below (Fig. S1) depicted the distances that $d$ and $Q$ measure.

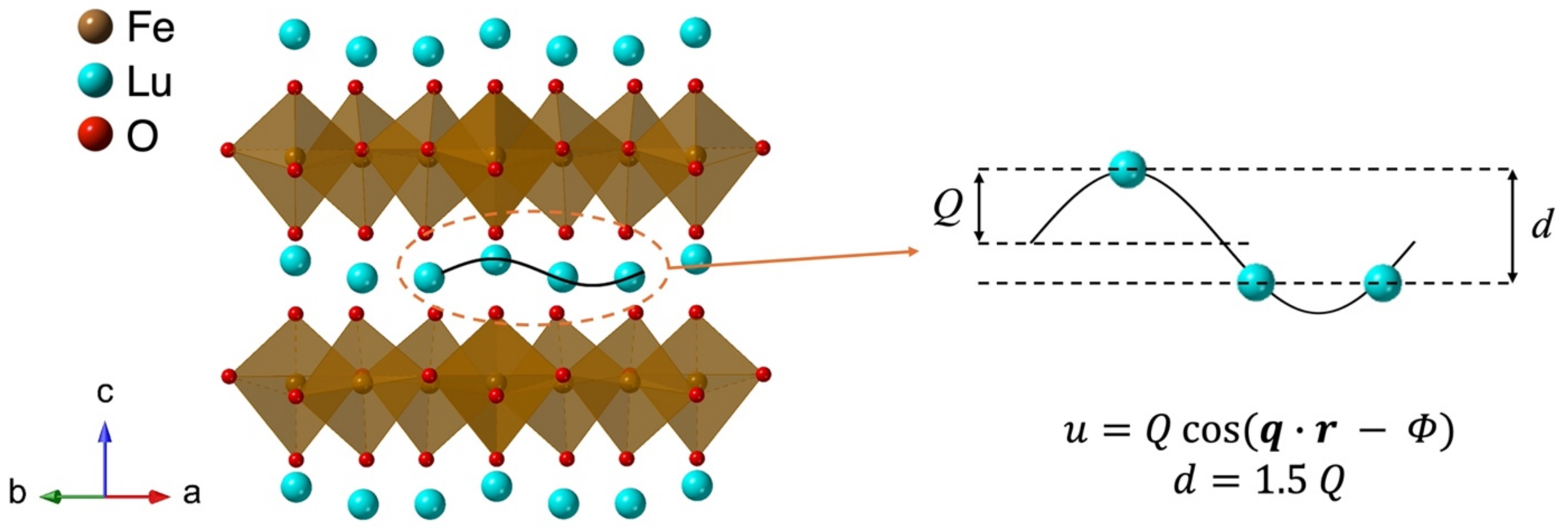

**Figure S1 | Schematic atomic structure of *h*-$LuFeO_3$ and the two quantitative descriptors of the extent of lattice trimerization (*Q* and *d*).**

From the established theory (Ref. 5), “a non-zero trimer distortion induces a non-zero polarization,” and there is a reliable relationship between $Q$, $d$ and $P$ in *h*-$LuFeO_3$. Extended Data Figure 2 in Ref. 6 offers the quantitative relationship between electric polarization $P$ and Lu displacement $d$. According to the charted values in Extended Data Figure 2 in Ref. 6, we can map the polarization values of our *h*-$LuFeO_3$ films with varying thicknesses that are grown on both SGMZ and sapphire substrates based on the measured $Q$ and $d$ values, as shown in Table S1 below. This supports the presence of net polarization down to 0.75-unit-cell thickness. Table S2 lists the electric polarization ($P$) values of bulk *h*-$LuFeO_3$ samples in various morphologies (e.g., film, powder) and measured using different methods. The electric polarization ($P$) values of these samples are found to range from 6.5 $\mu C/cm^2$ to 8.4 $\mu C/cm^2$. All the samples presented in our manuscript exhibit electric polarization ($P$) values within this range, reinforcing our claim that the ultrathin *h*-$LuFeO_3$ films in our study retain their full electric polarization without any noticeable diminishment.

**Table S1**

| **Samples from this manuscript (all measured at 300 K)** | | $Q_{\text{measured}}$ | $1.5\ Q_{\text{measured}}$ | $d_{\text{measured}}$ | ***P*** **(mapped from the chart in Ref. 6 using** $1.5\ Q_{measured}$**)** |
|---|---|---|---|---|---|
| 0.75-unit-cell-thick $h$-$LuFeO_3$ film on the engineered epitaxial template on an SGMZ(0001) substrate | | 25±1 pm | 37.5±1.5 pm | 37±6 pm | 6.4±0.3 μC/cm$^2$ |
| 0.75-unit-cell-thick $h$-$LuFeO_3$ film on the engineered epitaxial template on an $\alpha$-$Al_2O_3$(0001) substrate | | 28±1 pm | 42±1.5 pm | 41±5 pm | 7.4±0.3 μC/cm$^2$ |
| 2.25-unit-cell-thick $h$-$LuFeO_3$ film on the engineered epitaxial template on an SGMZ(0001) substrate | 1st Lu layer (bottom) | 30±2 pm | 45±3 pm | 41±7 pm | 8.1±0.7 μC/cm$^2$ |
| | 2nd Lu layer | 30±2 pm | 45±3 pm | 42±4 pm | 8.1±0.7 μC/cm$^2$ |
| | 3rd Lu layer | 31±3 pm | 46.5±4.5 pm | 46±3 pm | 8.4±1.0 μC/cm$^2$ |
| | 4th Lu layer (top) | 34±2 pm | 51±3 pm | 52±7 pm | 9.4±0.7 μC/cm$^2$ |

**Table S2**

| **Reference Source** | **Sample Morphology** | ***Q*** **or** ***d*** | **Measurement Method (at 300 K)** | ***P*** **(mapped from the chart in Ref. 6)** |
|---|---|---|---|---|
| Ref. 7 | 60 nm *h*-$LuFeO_3$ film/Pt(111) bottom electrode/ $\alpha$-$Al_2O_3$(0001) substrate | N/A | Polarization-field (P−E) hysteresis loops at 300K | ~6.5 μC/cm$^2$ |
| Extended Data Figure 5 in the manuscript | *h*-$LuFeO_3$ film/Ir(111) bottom electrode/YSZ(111) substrate | $Q$ = 31 pm; $d$ = 1.5$Q$ = 46.5 pm | STEM | ~8.4 μC/cm$^2$ |
| Ref. 8 | Phase-pure powder | $d$ = 45.5 pm | Powder XRD and crystal structure analysis | ~8.2 μC/cm$^2$ |
| Ref. 9 | Phase-pure powder | $d$ = 41.0 pm | Powder XRD and crystal structure analysis | ~7.2 μC/cm$^2$ |

**Section 3 — X-ray linear dichroism (XLD) measurements corroborating the ferroelectric order in ultrathin *h*-$LuFeO_3$ films.**

To corroborate the ferroelectric distortion seen by STEM in our ultrathin *h*-$LuFeO_3$ films, we performed XLD measurements on *h*-$LuFeO_3$ films with varying thicknesses and grown on two different substrates — SGMZ(0001) and $\alpha$-$Al_2O_3$(0001). The room-temperature XLD is shown in Fig. S2. For the 0.75-unit-cell (1 monolayer) *h*-$LuFeO_3$ film, the dichroic signal is sound and has the same signature shape as bulk *h*-$LuFeO_3$, which increases in strength with thickness. In *h*-$LuFeO_3$, the only asymmetries that contribute to the dichroic signal at room temperature are the ferroelectric distortions, indicating that even by a thickness of 0.75-unit cell there are ferroelectric structural distortions in the film, in agreement with the STEM analysis.

**XLD Method**: Measurements of the XLD spectroscopy were taken at the Advanced Light Source Beamline 4.0.2 at Lawrence Berkeley National Laboratory.

| Sample information | X-ray linear dichroism (XLD) |
|---|---|
| **0.75-unit-cell-thick *h*-$LuFeO_3$ film/ the engineered epitaxial template/ SGMZ(0001) substrate** | 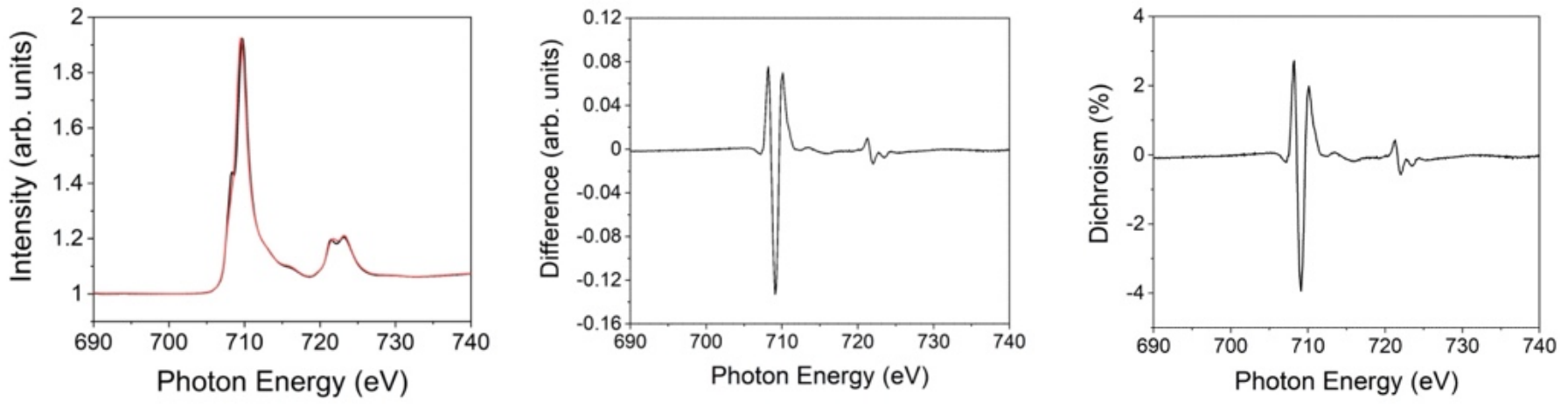 |
| **2.25-unit-cell-thick *h*-$LuFeO_3$ film/ the engineered epitaxial template/ SGMZ(0001) substrate** | 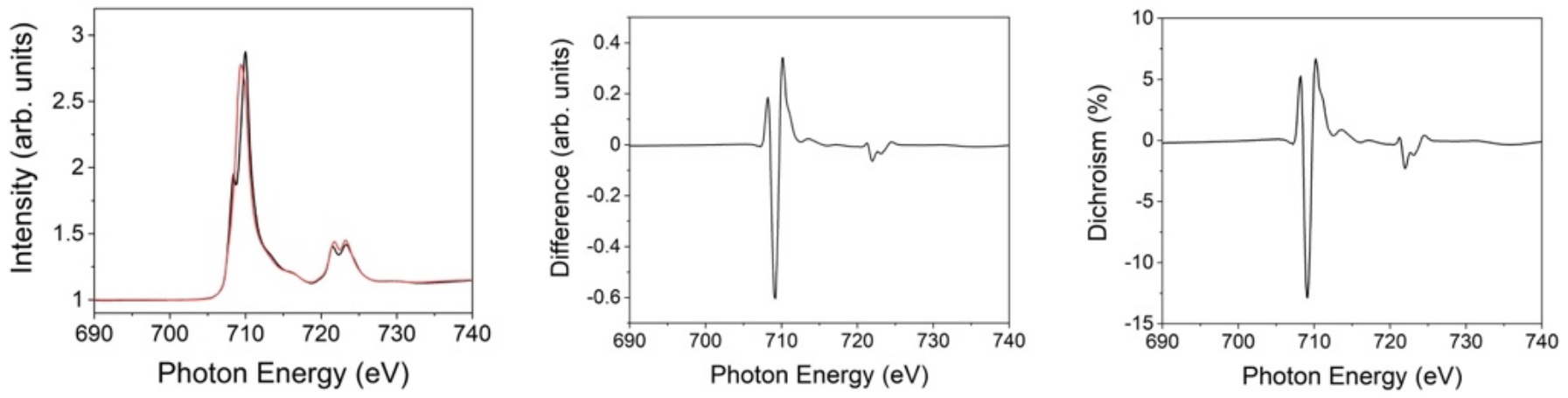 |
| **85.25-unit-cell-thick *h*-$LuFeO_3$ film/ the engineered epitaxial template/ SGMZ(0001) substrate** | 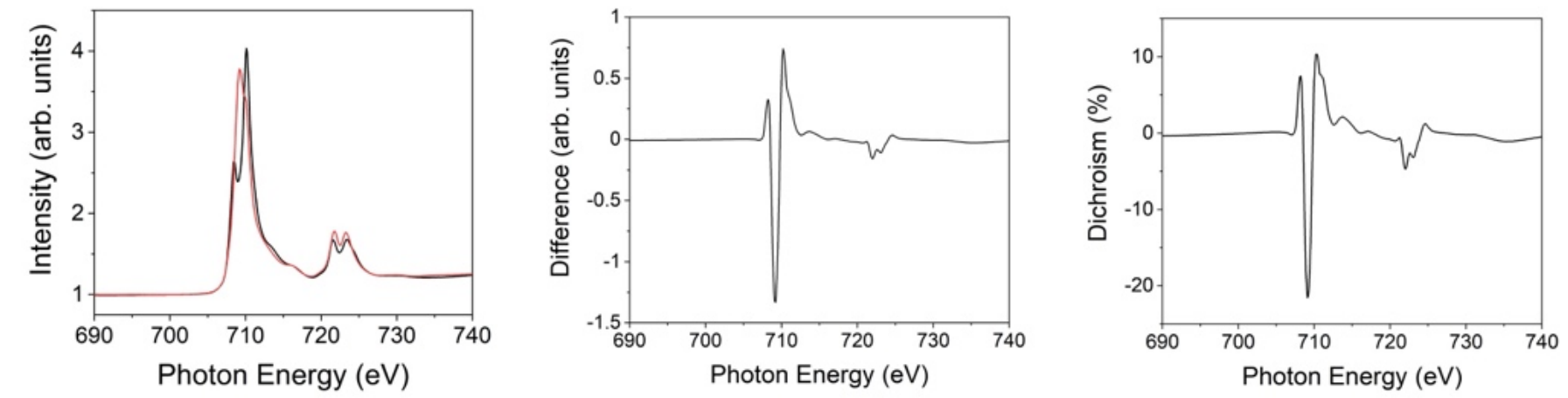 |
| **0.75-unit-cell-thick *h*-$LuFeO_3$ film/ the engineered epitaxial template/ *α*-$Al_2O_3$(0001) substrate** | 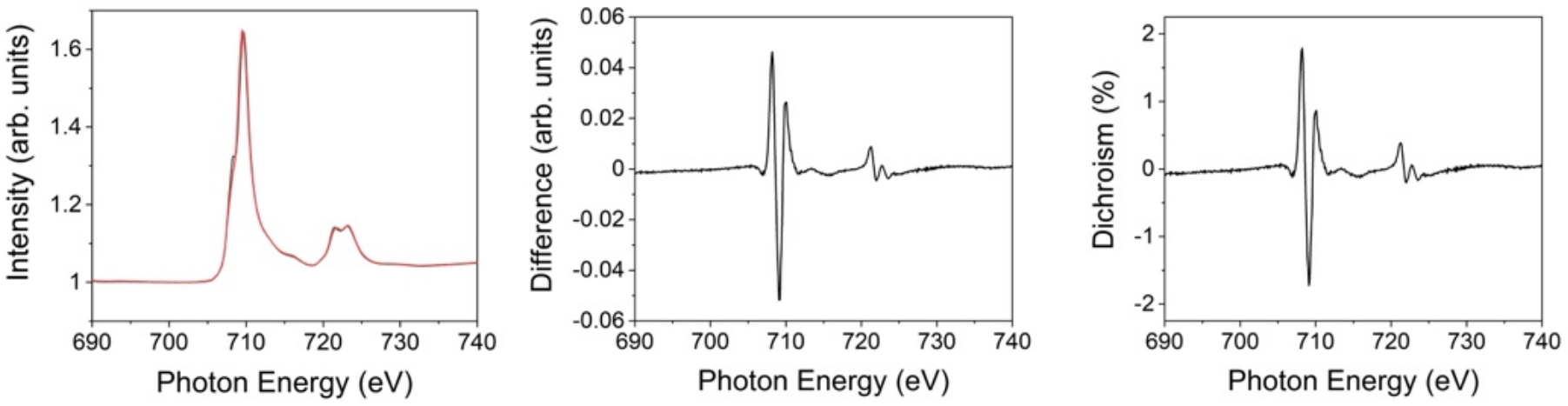 |
| **1.75-unit-cell-thick *h*-$LuFeO_3$ film/ the engineered epitaxial template/ *α*-$Al_2O_3$(0001) substrate** | 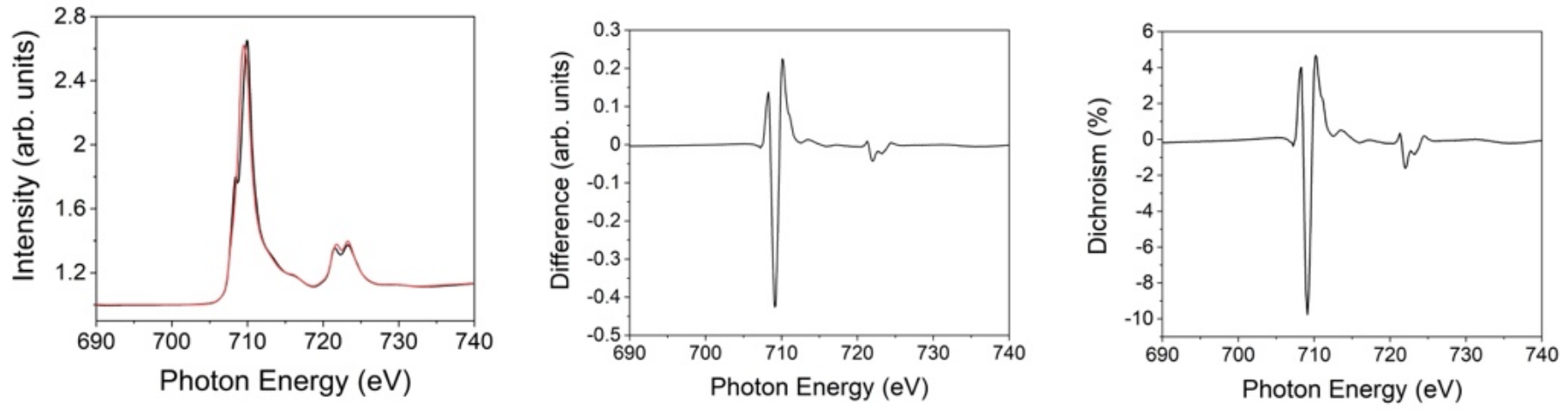 |
| **3.25-unit-cell-thick *h*-$LuFeO_3$ film/ the engineered epitaxial template/ *α*-$Al_2O_3$(0001) substrate** | 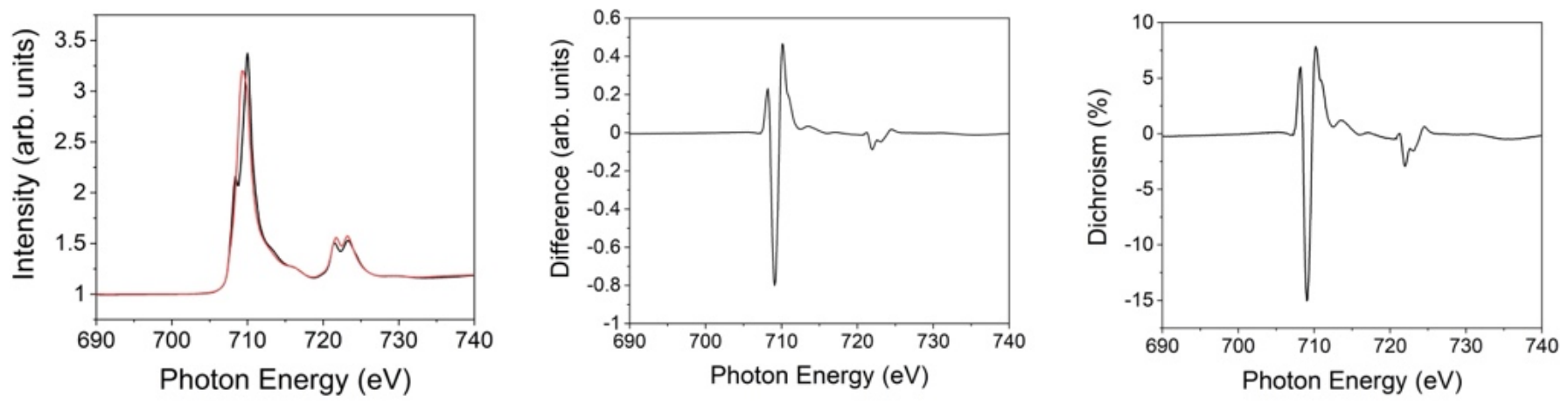 |

**Figure S2 | Room-temperature X-ray linear dichroism (XLD) measurements on *h*-$LuFeO_3$ films with varying thicknesses and grown on two different substrates — SGMZ(0001) and $\alpha$-$Al_2O_3$(0001).**

**Section 4 — Second harmonic generation measurements supporting that the one-third-order streaks in RHEED corresponding to the ferroelectric phase are not from some surface effect.**

Temperature-dependent second harmonic generation measurements are performed on a 35 nm thick and on a 100 nm thick *h*-$LuFeO_3$ film grown on a $SrCo_2Ru_4O_{11}$ bottom electrode on an SGMZ substrate. Figure S3a shows the geometry of the measurement setup. An ideal incident laser power window exists where the SHG signals are only from the *h*-$LuFeO_3$ films and not from the underlying epitaxial template, as shown in Fig. S3b. Due to the possibility of nanoscale domain variants with opposite polarization directions in these films leading to out-of-phase second harmonic fields from different domains, the relative SHG signal strengths cannot be compared for the two *h*-$LuFeO_3$ films with different thicknesses. In Fig. S3c,d, SHG polarimetry of both films at either room temperature or high temperature (1010 K) fits nicely with the non-centrosymmetric 6*mm* point group, which the ferroelectric phase of *h*-$LuFeO_3$ is expected to have. As temperature increases from 300 K to 1010 K, as shown in Fig. S3e, the SHG intensity of the 35 nm thick and the 100 nm thick *h*-$LuFeO_3$ films both decrease, but do not reduce to the noise level at 1010 K, demonstrating that the Curie temperatures of both films are higher than 1010 K. At a growth temperature of ~1000 K, the one-third-order streaks in RHEED are observed at the end of the growth of both films (Fig. S4a,b), corroborating the SHG result that the Curie temperatures of both films are higher than ~1000 K. In addition to Refs. 13,31 in the main text, this further supports that the onset of ferroelectricity in $h$-$LuFeO_3$ can be observed *in situ* during growth using RHEED.

Detailed methods used for measuring and fitting the SHG polarimetry are described as follows. SHG polarimetry and temperature dependent measurements were performed with a femtosecond pulsed $\lambda$ = 800 nm fundamental light from a regeneratively amplified Spectra-Physics Solstice Ace Ti:Sapphire laser system (1 kHz, 100 fs). The fundamental light with $\lambda$ = 800 nm is focused to ~ 30 µm beam diameter. The $p$ and $s$-polarized SHG intensities are spectrally filtered and measured by a photomultiplier tube through a lock-in amplifier (SR830).

A schematic of the second harmonic generation setup is shown in Fig. S3a. The lab coordinates (X, Y, Z) are related to the direction of the incoming beam with X || $p$ polarization and Y || $s$ polarization ($p$ and $s$ are defined accordingly in Fig. S3a).

In lab coordinates (X, Y, Z), the electric field of the incident beam can be written as $(E_o \cos\varphi,\ E_o sin\ \varphi, 0)$ where $\varphi$ is the polarization rotation angle introduced by the half wave plate as shown in Fig. S3a. For an incidence angle $\theta$ on the sample (see Fig. S3a), the electric field in the crystal axes coordinates can be expressed as $(E_o\cos\ \varphi\cos\ \theta, E_o\sin\ \varphi, -E_o\cos\ \varphi\sin\ \varphi)$. The induced nonlinear polarization, $P^{2\omega}$, is related to the incident electric field through the nonlinear susceptibility tensor, $d_{ijk}$ through the following equation:

$$P_i^{2\omega} \propto d_{ijk} E_j^{\omega} E_k^{\omega} \tag{1}$$

The proportionality constants depend on incident beam fluence, Fresnel's coefficients at the film-air and film-substrate interfaces and the thickness of the films.

For point group 6*mm* with the 6-fold axis lying along Z – directions, the nonlinear susceptibility tensor can be written in Voigt notation as:

$$\begin{pmatrix} 0 & 0 & 0 & 0 & d_{15} & 0 \\ 0 & 0 & 0 & d_{15} & 0 & 0 \\ d_{31} & d_{31} & d_{33} & 0 & 0 & 0 \end{pmatrix}$$

From this, using equation (1) we can calculate $p$ and $s$ polarized SHG intensities for $\theta = 45°$ angle of incidence as:

$$I_p^{2\omega} \propto \left(P_p^{2\omega}\right)^2 \propto ((2d_{15} - d_{31} - d_{33})cos\,[\varphi]^2 - 2d_{31}sin\,[\varphi]^2)^2 \qquad (2)$$

$$I_s^{2\omega} \propto (P_s^{2\omega})^2 \propto d_{15}^2 Sin\,[2\varphi]^2$$

Equation 2 is used to fit the SHG polarimetry from the 35 nm and 100 nm thick *h*-$LuFeO_3$ films both at 300 K and 1010 K (Fig. S3c-d).

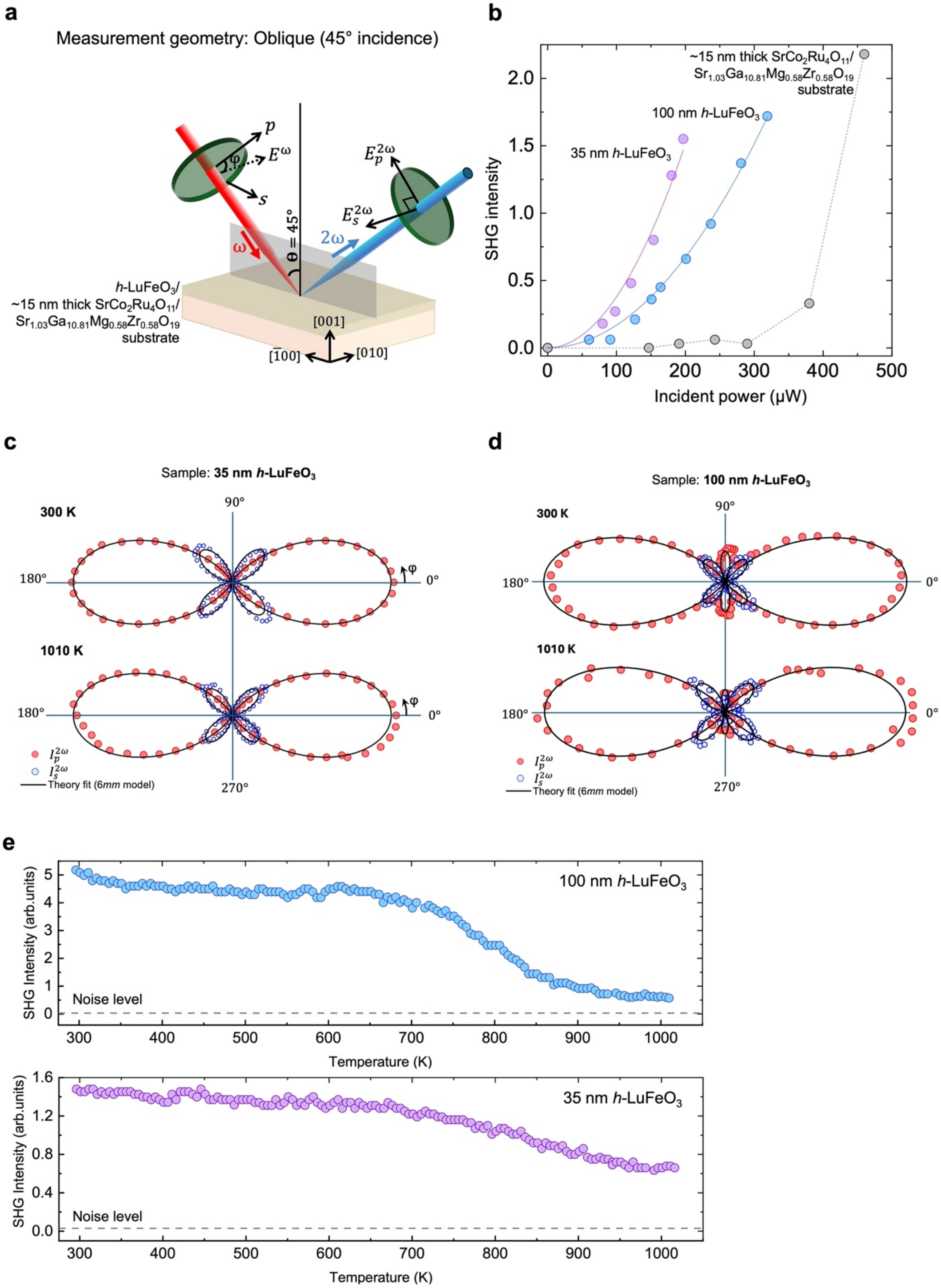

**Figure S3 | Temperature-dependent second harmonic generation (SHG) measurements on 35 nm and 100 nm thick *h*-$LuFeO_3$ samples grown on $SrCo_2Ru_4O_{11}$ bottom electrodes on SGMZ substrates. a,** SHG measurement geometry. **b,** Incident laser-power-dependent SHG measurement of *h*-$LuFeO_3$ films (grown on $SrCo_2Ru_4O_{11}$ bottom electrodes on SGMZ substrates) vs. bare $SrCo_2Ru_4O_{11}$ bottom electrodes on SGMZ substrate. **c,** Room- and high-temperature polarimetry of the 35 nm thick *h*-$LuFeO_3$ film grown on a $SrCo_2Ru_4O_{11}$ bottom electrode on a SGMZ substrate, both fitting to the 6*mm* point group. **d,** Room- and high-temperature polarimetry of the 100 nm thick *h*-$LuFeO_3$ film grown on a $SrCo_2Ru_4O_{11}$ bottom electrode on a SGMZ substrate, both fitting to the 6*mm* point group. **e,** Temperature dependence of the SHG intensity of the 35 nm thick and the 100 nm thick *h*-$LuFeO_3$ films grown on $SrCo_2Ru_4O_{11}$ bottom electrodes on SGMZ substrates.

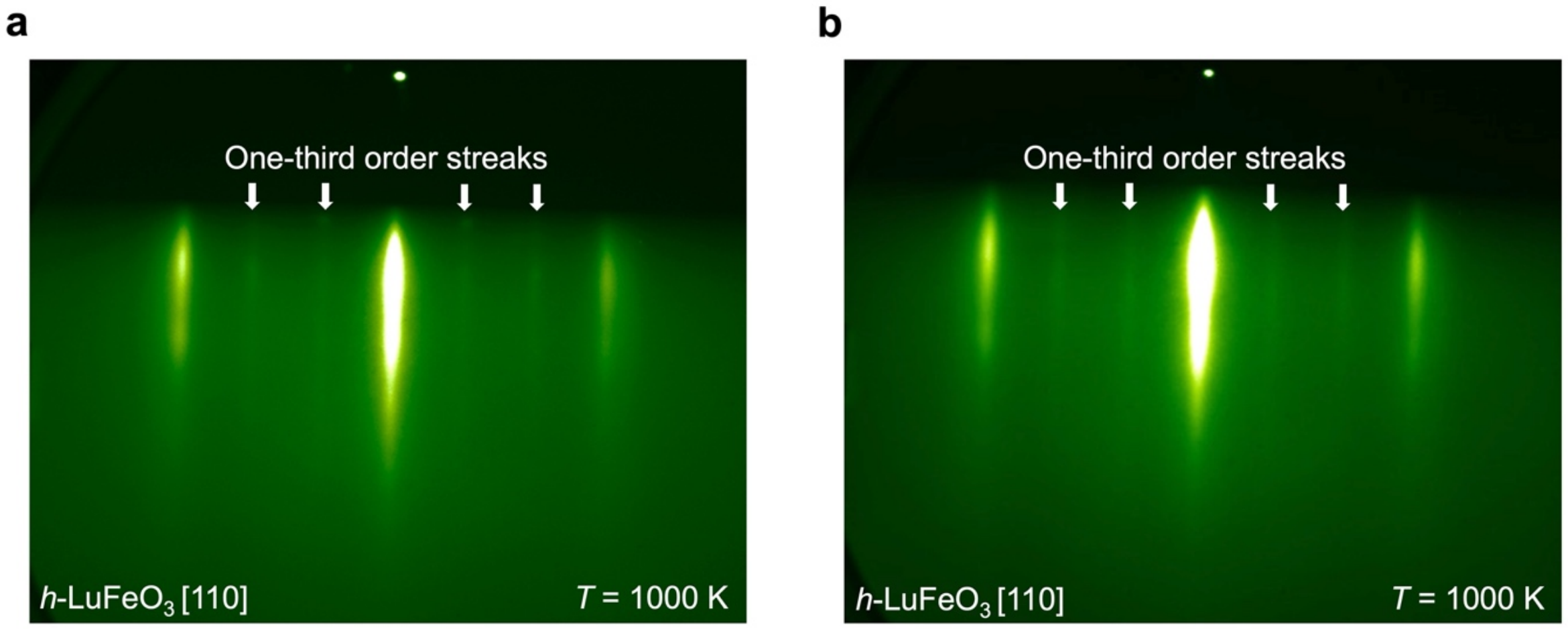

**Figure S4 | *In-situ* RHEED at the end of the growth of 35 nm and 100 nm thick *h*-$LuFeO_3$ samples at growth temperature. a,** One-third-order streaks in the *in-situ* RHEED image at the end of the growth of the 35 nm thick *h*-$LuFeO_3$ film showing the structural in-plane tripling of the unit cell leading to ferroelectricity. This image was taken at 1000 K, measured by an optical

pyrometer. **b,** One-third-order streaks in the *in-situ* RHEED image at the end of the growth of the 100 nm thick *h*-$LuFeO_3$ film showing the structural in-plane tripling of the unit cell leading to ferroelectricity. This image was taken at 1000 K, measured by an optical pyrometer.

**Section 5 — PFM measurements of polarization switching and electrical measurements of switching current.**

We conducted ferroelectric switching studies with piezoresponse force microscopy (PFM) measurements. Our results show that we can electrically switch films with *h*-$LuFeO_3$ thicknesses of 0.75-unit cells (Main text Fig. 3), 2.25-unit cells (Fig. S5), 3.25-unit cells (Fig. S6), and 12.25-unit cells (Fig. S7). Furthermore, the polarization of these films can be mechanically switched from upward to downward (the downward polarization state remains intact under the mechanical load), avoiding voltage related issues such as charge injection (Ref. 10). The similarity of the PFM responses after electrical and mechanical poling in these films further implies ferroelectric switching. PFM switching spectroscopies (Figs. S5c, S6c, S7c) reveal conventional hysteresis loops corroborating ferroelectric switching behavior.

Electrical measurements of the 12.25-unit-cell-thick *h*-$LuFeO_3$ capacitor structure with a graphene flake used as a top electrode (Fig. S7d) produce a clear switching current peak on the negative voltage side, while strong leakage did not allow observation of similar results for the positive voltage.

Our ability to switch films at atomic-scale thicknesses represents the thinnest improper ferroelectric switching observed to date.

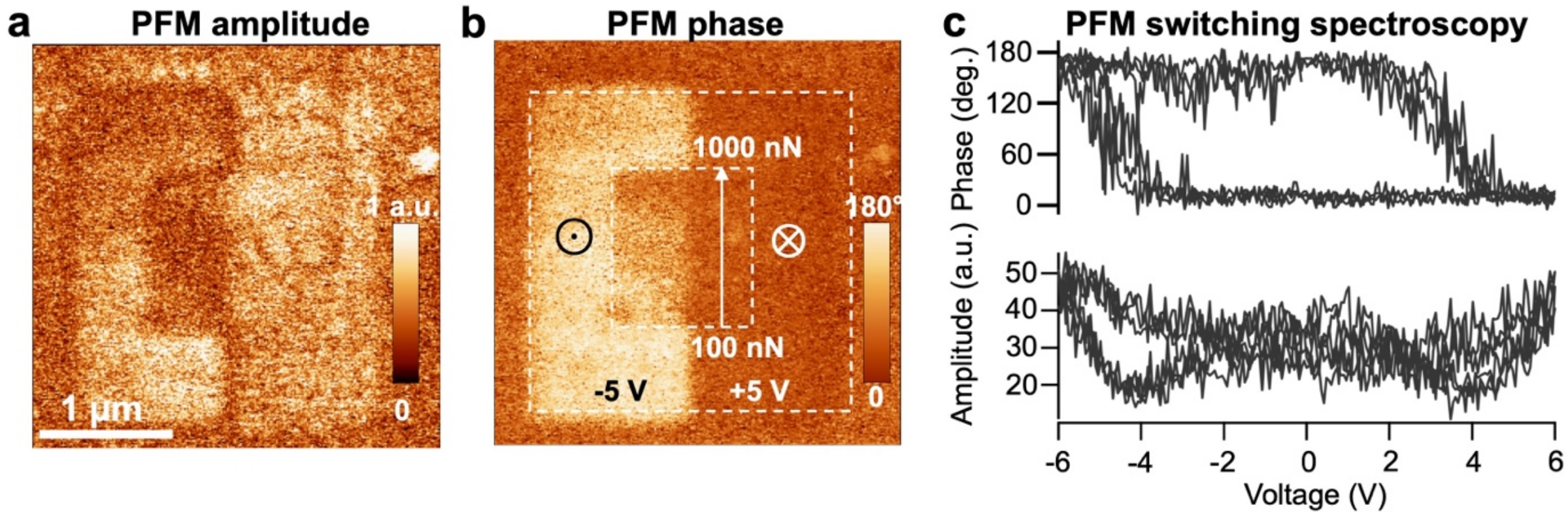

**Figure S5 | Polarization switching confirmed by PFM measurements in the 2.25-unit-cell-thick *h*-$LuFeO_3$ film on the engineered epitaxial template on an SGMZ substrate. a-b,** PFM amplitude and phase images obtained after electrical writing of ±5 V with the PFM tip in a 2×2 $\mu m^2$ region (left side, -5 V, and right side, +5 V), followed by mechanical writing in a 1×1 $\mu m^2$ center square region with loading force from 100 nN to 1000 nN from bottom to top on the free surface of the *h*-$LuFeO_3$ film. **c,** Representative PFM switching spectroscopy obtained at a fixed location from the *h*-$LuFeO_3$ film showing conventional butterfly-like strain loop, with coercive voltage around 4 V.

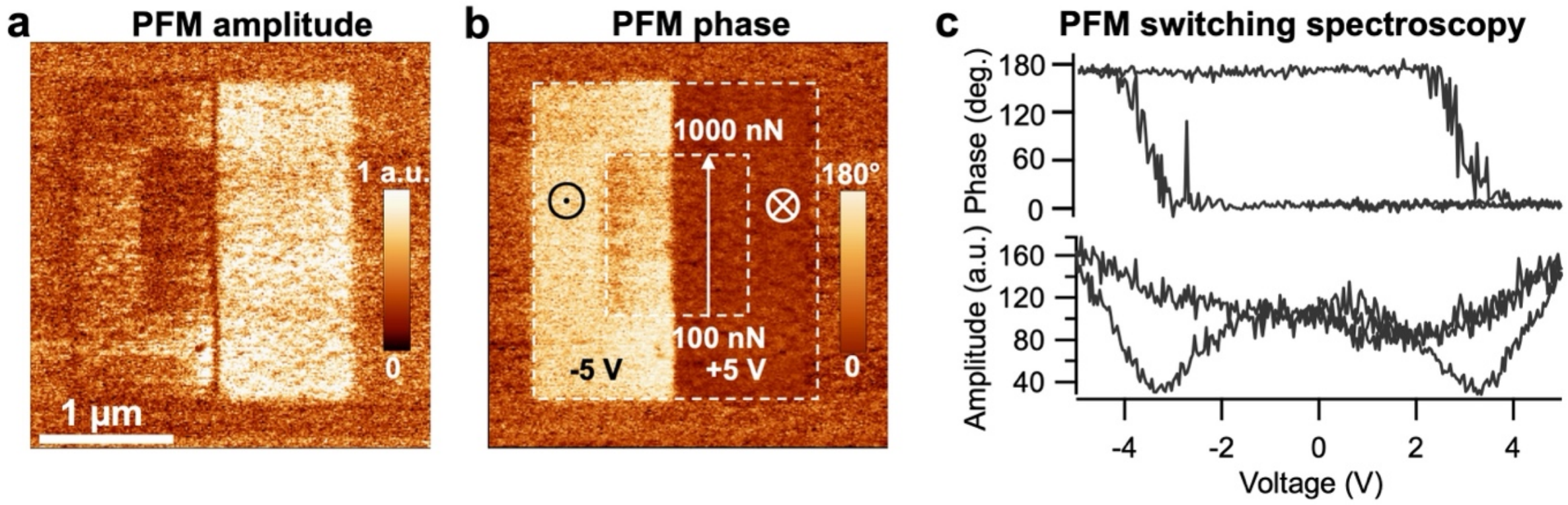

**Figure S6 | Polarization switching confirmed by PFM measurements in the 3.25-unit-cell-thick *h*-$LuFeO_3$ film on the engineered epitaxial template on a *c*-plane sapphire substrate.**

**a-b,** PFM amplitude and phase images obtained after electrical writing of ±5 V with the PFM tip in a 2×2 μm$^2$ region (left side, -5 V, and right side, +5 V), followed by mechanical writing in a 1×1 μm$^2$ center square region with loading force from 100 nN to 1000 nN from bottom to top on the free surface of the *h*-$LuFeO_3$ film. **c,** Representative PFM switching spectroscopy obtained at a fixed location from the *h*-$LuFeO_3$ film showing conventional butterfly-like strain loop, with coercive voltage around 3.5 V.

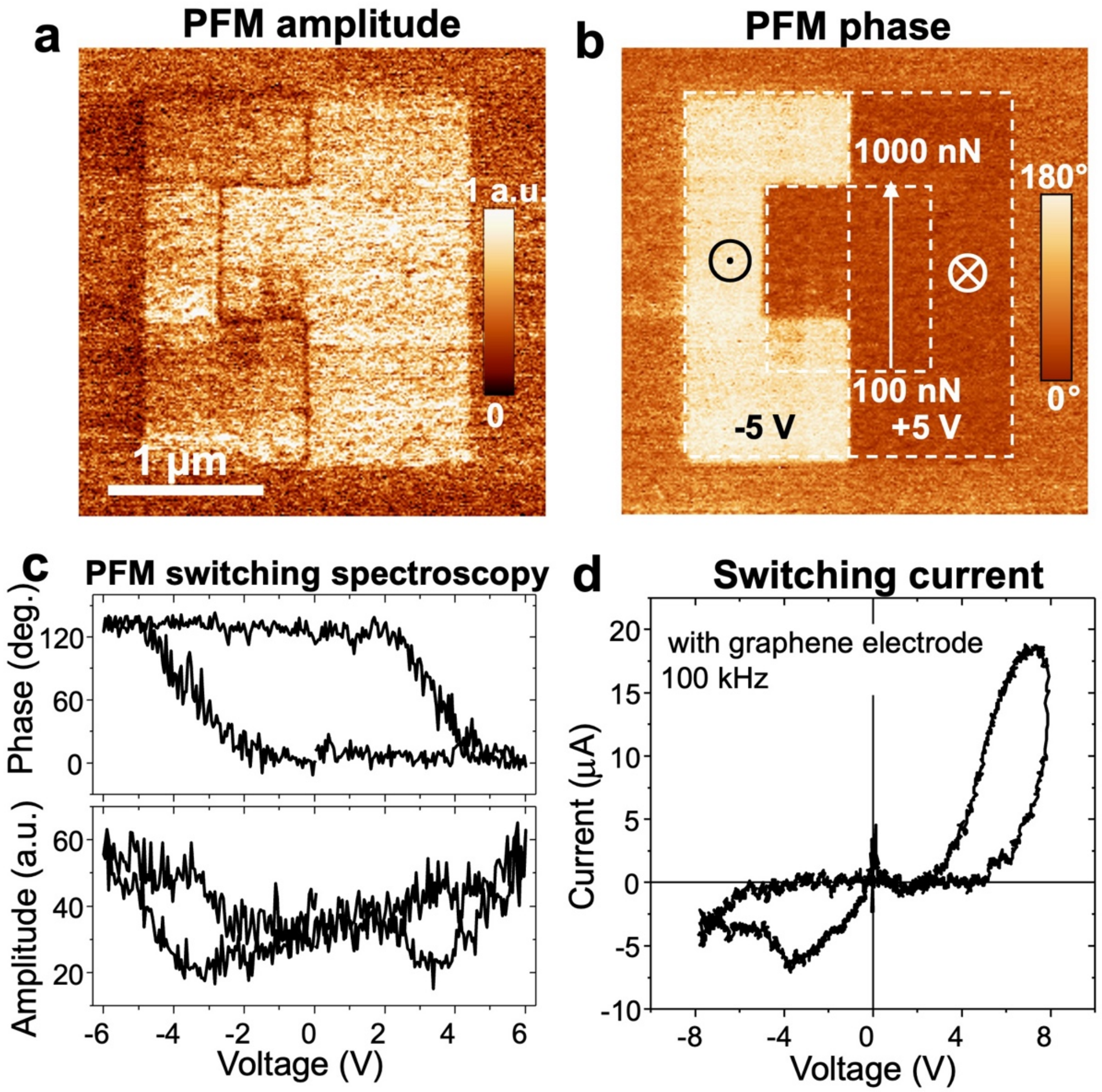

**Figure S7 | PFM measurements of polarization switching and electrical measurements of switching current in the 12.25-unit-cell-thick *h*-$LuFeO_3$ film on the engineered epitaxial template on a *c*-plane sapphire substrate. a-b,** PFM amplitude and phase images obtained after electrical writing of ±5 V with the PFM tip in a 2×2 $\mu m^2$ region (left side, -5 V, and right side, +5 V), followed by mechanical writing in a 1×1 $\mu m^2$ center square region with loading force from 100 nN to 1000 nN from bottom to top side on the free surface of the *h*-$LuFeO_3$ film. **c,** Representative PFM switching spectroscopy obtained from the *h*-$LuFeO_3$ film showing conventional butterfly-like strain loop, with coercive voltage around 3.5 V. **d,** Switching current measured at 100 kHz from a capacitor with 210 $\mu m^2$ graphene top electrode. Note that the leakage current at the negative voltages has been fully compensated, while some leakage at the positive voltage side cannot be fully compensated.